\documentclass[11pt,english]{article}
\usepackage[T1]{fontenc}
\usepackage{a4}
\usepackage{graphicx}
\usepackage{setspace}
\IfFileExists{url.sty}{\usepackage{url}}
                      {\newcommand{\url}{\texttt}}

\makeatletter

\providecommand{\LyX}{L\kern-.1667em\lower.25em\hbox{Y}\kern-.125emX\@}

\usepackage{mathptmx}
\usepackage{times}

\usepackage{babel}
\makeatother
\begin{document}

\title{Introduction to Quantum Computation%
\footnote{Based on slightly extended and expanded version of lectures given
at the Bose Institute, Kolkata and Saha Institute Of Nuclear Physics,
Kolkata, during November-December 2002%
}}

\author{Ashok Chatterjee%
\footnote{e-mail: ashok@theory.saha.ernet.in%
} }

\date{\emph{Theory Group, Saha Institute Of Nuclear Physics }\\
\emph{Kolkata 700 064, India}}

\maketitle

\section{Introduction}

Computation basically is digital data processing. This requires the
use of a computer which processes the data following certain set of
instructions called a programme. Examples are, numerical data being
processed by the executable version of a FORTRAN or C programme, text
being edited by a word processor and a visual image being rendered
by a graphics application.

The input, intermediate and output data are internally expressed in
terms of certain basic units called bits. Each bit has two possible
values $0$ and $1$ and a string of such values corresponds to the
binary representation of a number just like the more familiar decimal
representation. The advantage of using the binary representation in
a computer is that, it is relatively easy to construct devices that
possess two clearly distinguishable states that may be used to represent
the bit values. Examples are high and low voltage states of a capacitor
and the two stable states of a flip-flop circuit.

The devices used to represent bits behave essentially as classical
systems, even though they may be inherently dependent on quantum phenomena
for their operation. For example, a transistor used in a flip-flop
circuit works on the basis of the semi-conducting properties of certain
materials. This stems from the quantum mechanical energy band structure
of electrons in those materials. However, because of the large number
of electrons involved, quantum effects due to them add up incoherently
to produce say, a current or voltage that behaves classically. The
flip-flop or the capacitor, consequently exists in one of the two
possible stable states and not in any arbitrary mixture of them. Thus
at any time, a processor using such devices to represent and store
bits, can only process a particular set of data. In order to process
several sets of data concurrently, one has to use several such processors
and run parallel data channels through them. This is ordinary parallel
processing.

There are on the other hand intrinsically quantum systems which have
two orthogonal basis states that may be used to represent the two
values of a bit. Examples are the up and down states of a spin half
object, the two orthogonal polarization states of a photon and two
non-degenerate energy eigenstates of an atom. Let us represent the
two states corresponding to bit values $0$ and $1$ by $\left|0\right\rangle $
and $\left|1\right\rangle $ respectively. However these are not the
only possible states for a quantum mechanical bit or \emph{qubit}
as they are called. Because of the superposition principle in quantum
mechanics, an arbitrary linear combination $\alpha \left|0\right\rangle +\beta \left|1\right\rangle $
with complex coefficients $\alpha $ and $\beta $ is also a possible
state. A qubit in such a state, in a sense, carries the two possible
bit values simultaneously. A device based on such qubits possesses
the potential for massive parallelism that may be harnessed to construct
quantum computers which are immensely more powerful than their classical
counterparts. We expect such a computer to be particularly useful
in simulating efficiently quantum systems such as an atom, which is
a task for which classical computers are generally extremely inadequate.

However, the catch is that a qubit, when measured at the end of a
computation always collapses to one or the other basic state, yielding
a value which is either $0$ or $1$. Thus, even though it may be
possible for a quantum computer to carry out a large number of computations
on different sets of data parallely, at the end of the day we obtain
the result for just one of the sets. Notwithstanding this difficulty,
it is possible with clever design of quantum algorithms and quantum
devices to implement them, to use quantum computers to solve certain
problems that are very hard to solve otherwise.

\section{Classical Gates}

As we discussed earlier, classical computation consists of processing
or transformation of data represented by classical bits. The elementary
units that process classical bits are called gates. Processors used
in modern electronic computers use tens and hundreds of millions such
gates. The design is modular. So we don't have to understand how these
gates really work. It will be enough to treat them as little black
boxes with specified inputs and the corresponding outputs. In that
case we do not have to worry when the internal design of a gate changes,
as long as the external function remains the same.
\newpage

The classical gates are classified according to the number of inputs.

\begin{itemize}
\item \textbf{Single-input gates:}
\end{itemize}
(A) The \textbf{NOT} gate: This simply switches the value of the input
bit from $0$ to $1$ and \emph{vice-versa. }

\begin{center}\includegraphics{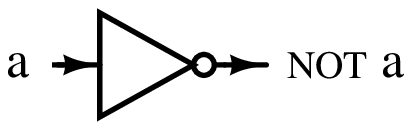} \end{center}

\begin{flushleft}Mathematically: $NOT\: \: a=1\oplus a$ , where$\: \: \oplus \: \: $indicates
addition mod $2$.\end{flushleft}

\begin{flushleft}(B) The \textbf{FANOUT} (Copy) gate: This is simply
a wire carrying the input bit that branches out into two others carrying
the same bit. \end{flushleft}

\begin{center}\includegraphics{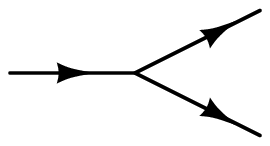}\end{center}

\begin{flushleft}(C) The \textbf{ERASE} gate: This simply erases or
resets to 0 the input bit.\end{flushleft}

\vspace{0.375cm}
\begin{center}\includegraphics{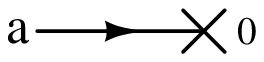}\end{center}
\vspace{0.375cm}

\begin{itemize}
\item \textbf{Two-input gates:}
\end{itemize}
(A) The \textbf{AND} gate:

\vspace{0.3cm}
\noindent \begin{center}\includegraphics{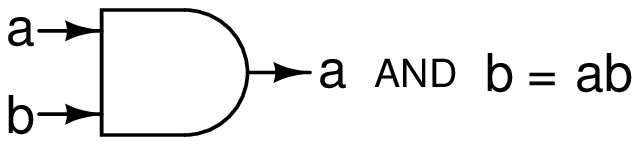} \end{center}

\noindent \begin{flushleft}This produces a single output bit from
two input bits. The output is 1 only if both inputs are 1 and is 0
otherwise. Mathematically $a\: AND\: b=ab$.\end{flushleft}

\newpage
\begin{flushleft}(B) The \textbf{OR} gate:\end{flushleft}

\vspace{0.3cm}
\begin{center}\includegraphics{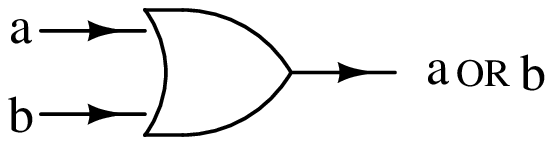} \end{center}

\begin{flushleft}The output in this case is 0 only if both the inputs
are 0 and is 1 otherwise.\end{flushleft}

\noindent \begin{flushleft}(C) The \textbf{XOR} gate:\end{flushleft}

\vspace{0.3cm}
\begin{center}\includegraphics{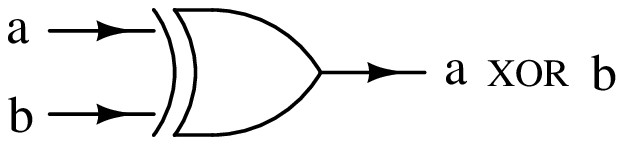} \end{center}

\begin{flushleft}Mathematically $a\: XOR\: b=a\oplus b$ . The output
in this case is 1 only if one of the inputs is 1 and the other is
0 . Otherwise the output is 0 .\end{flushleft}

The two-input gates are conveniently described in terms of truth tables
which display the outputs for various possible inputs. The truth tables
of AND, OR and XOR \textbf{}gates are shown together below.

\vspace{0.375cm}
\begin{center}\begin{tabular}{|c|c|c|c|c|}
\hline 
\textbf{a}&
\textbf{b}&
\textbf{AND}&
\textbf{OR}&
\textbf{XOR}\\
\hline
\hline 
0&
0&
0&
0&
0\\
\hline
0&
1&
0&
1&
1\\
\hline
1&
0&
0&
1&
1\\
\hline
1&
1&
1&
1&
0\\
\hline
\end{tabular}\end{center}
\vspace{0.375cm}

\begin{flushleft}From the table we easily verify\end{flushleft}

\begin{flushleft}\[
a\: OR\: b=(a\: AND\: b)\: XOR\: (a\: XOR\: b)\]
 \end{flushleft}

\begin{flushleft}Thus the OR \textbf{}gate may be constructed by combining
the \textbf{}AND \textbf{}and the \textbf{}XOR \textbf{}gates.\end{flushleft}

\newpage
\begin{flushleft}Next we may combine the NOT \textbf{}gate with either
the AND \textbf{}gate or the OR \textbf{}gate to obtain two more gates.\end{flushleft}

\begin{flushleft}(D) The \textbf{NAND} gate:\end{flushleft}

\begin{center}\includegraphics{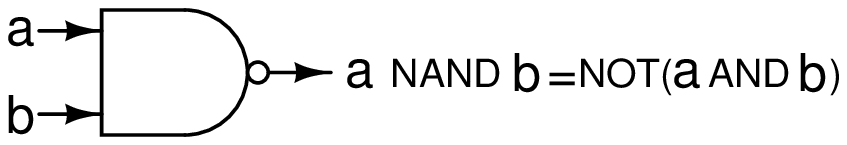}\end{center}
\medskip{}

\begin{flushleft}(E) The \textbf{NOR} gate:\end{flushleft}

\vspace{0.3cm}
\begin{center}\includegraphics{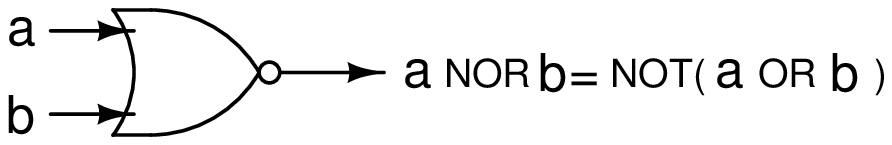}\end{center}

It turns out that any classical computation can be implemented by
a circuit constructed out of the above set of single-input and two-input
gates. More surprisingly, it happens that we do not even need all
the above gates. For example, the other gates can be made out of ERASE,
FANOUT and NAND \textbf{}gates which therefore form a minimal universal
set.

\section{Reversible Computation}

While the inputs of NOT \textbf{}and FANOUT \textbf{}gates may be
reconstructed from the outputs, the same is not true for other classical
gates. It is in this sense these other gates are not \emph{reversible.}
For each such gate the output contains one less bit than the input.
Since a classical bit has two possible states, the phase space is
reduced and there is a decrease in entropy which according to the
Boltzmann relation equals $k\ln 2$. According to the second law of
thermodynamics, this must be over-compensated by a corresponding increase
in the entropy of the surrounding. This is equivalent to heat released
to the environment at temperature $T$ , given by\[
\Delta Q\geq kT\: \ln \: 2\]

\begin{flushleft}This is exactly like the heat given off when one
molecule of an ideal gas is isothermally compressed to half the original
volume. Thus classical computation using \emph{irreversible} gates
inevitably generates heat. For present day computers this heat is
of course negligible compared to the heat generated by other dissipative
processes such as the flow of current through a resistance. As we
will see later, one important way in which the corresponding quantum
gates and computers based on them are different, is that they are
always reversible.\end{flushleft}

It is also possible to construct classical gates that are reversible.
The general idea is to copy some of the input bits to the output so
that the input bits may be reconstructed out of the result and the
extra output bits. An elegant implementation of this idea is  

\bigskip{}
The \textbf{Toffoli} gate:

\vspace{0.3cm}
\begin{center}\includegraphics{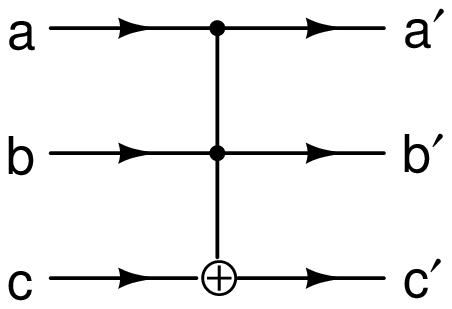}\end{center}
\vspace{0.3cm}

\begin{flushleft}where $a^{'}=a$ , $b^{'}=b$ and $c^{'}=c\oplus ab$.\end{flushleft}

\begin{flushleft}Since $a=a^{'}$ , $b=b^{'}$ and $c=c^{'}\oplus a^{'}b^{'}$
, the input bits may be reconstructed from the output bits simply
by running the gate in the reverse. Thus the Toffoli \textbf{}gate
is reversible.\end{flushleft}

\begin{flushleft}For $b=c=0$ , $c^{'}=0$ and the Toffoli gate acts
as a reversible ERASE gate. For $b=1\, $and $c=0$, $c^{'}=a$ and
it simulates a reversible FANOUT gate. For $b=1$ and $c=1$, $c^{'}=1\oplus a=NOT\: a$
and it is equivalent to a reversible NOT gate. For $b=1$, $c^{'}=c\oplus a=a\: \: XOR\: \: c$
and it acts as a reversible XOR gate. Finally for $c=0$, $c^{'}=ab=a\: \: AND\: \: b$
and it simulates a reversible AND. Thus the Toffoli gate may be used
to construct the reversible versions of all basic one input and two
input classical gates.\end{flushleft}

Most reversible gates use extra \emph{ancilla} bits (\emph{i.e.} auxiliary
bits that are set to standard states) at the input and produce extra
garbage bits (\emph{i.e.} bits other than those containing the results
of computation) at the output. These are in fact necessary to ensure
reversibility of those gates. However the garbage bits have to be
erased or reset at the end of the computation (or earlier) because
of the limitations on the available memory. One may worry that such
erasure would spoil the reversibility and generate heat. Fortunately
there is a way to erase the garbage without destroying reversibility.

To see how it works, let us consider a somewhat simple situation.
Suppose that the computation starts with the ancilla bits in the state
$0$ and the target bit in the state $x$. If some of the ancilla
bits are required to be in the state $1$, we can always arrange that
using NOT gates on those. Now a reversible computation produces the
result $r(x)$ and some garbage $g(x)$ . So it looks like 

\begin{flushleft}\[
(x,0,0)\rightarrow (x,r(x),g(x))\]
\end{flushleft}

\begin{flushleft}We may remove the last garbage bit reversibly by
simply reversing the computation (\emph{uncomputation) .} But that
removes the result $r(x)$ too. So we carry an extra ancilla bit again
in the state $0$ and reversibly copy the result of the computation
to that. So it looks like\[
(x,0,0,0)\rightarrow (x,r(x),g(x),0)\rightarrow (x,r(x),g(x),r(x))\]
\end{flushleft}

\begin{flushleft}Now we run the computation proper backwards. This,
of course, does not affect the fourth bit. So this final step is \[
(x,r(x),g(x),r(x))\rightarrow (x,0,0,r(x))\]
 \end{flushleft}

\begin{flushleft}We have succeeded in removing the garbage reversibly
without destroying the result!\end{flushleft}

\bigskip{}
\begin{flushleft}\textbf{Problem 1:} What reversible gate is simulated
by the Toffoli gate for $c=1$?\end{flushleft}
\bigskip{}

\bigskip{}
\begin{flushleft}\textbf{Problem 2:} Construct the truth table for
the Toffoli gate i.e. a table displaying the values of the output
bits $a^{'}$, $b^{'}$ and $c^{'}$ for all possible values of the
input bits $a$, $b$ and $c$.\end{flushleft}
\bigskip{}

\bigskip{}
\begin{flushleft}\textbf{Problem 3:} \emph{The Fredkin gate} is another
3-input 3-output gate just like the Toffoli gate. For this gate $c^{'}=c$.
If $c=0$ then $a^{'}=a$ and $b^{'}=b$ i.e. nothing happens. On
the other hand if $c=1$ then $a^{'}=b$ and $b^{'}=a$ i.e. the two
bits are swapped. Show that the Fredkin gate is reversible and explain
how it may be used to simulate a reversible AND gate. \end{flushleft}
\bigskip{}

\section{Quantum Gates}

The quantum gates transform qubits just like the way classical bits
are changed by the classical gates. However, this involves the time
evolution of a quantum system and according to the laws of quantum
mechanics this is described by a unitary operator. Thus to every quantum
gate corresponds a unitary operator $U$. So a quantum gate acts on
an arbitrary multi-qubit state $\left|\psi _{in}\right\rangle $ as\[
\left|\psi _{in}\right\rangle \rightarrow \left|\psi _{out}\right\rangle =U\left|\psi _{in}\right\rangle \]

\begin{flushleft}The input state may be reconstructed from the output
by\[
\left|\psi _{in}\right\rangle =U^{\dagger }\left|\psi _{out}\right\rangle \]
\end{flushleft}

\begin{flushleft}Thus the quantum gates are always reversible.\end{flushleft}

Because of linearity of the unitary operators, a quantum gate is described
completely by its action on a convenient basis. For a single qubit
the most convenient choice is the one that corresponds to the possible
results of measurement $0$ and $1$ \emph{i.e.} the states $\left|0\right\rangle $
and $\left|1\right\rangle $ . This is known as the \emph{computational
basis.} For multi-qubit states the computational basis is obtained
by tensoring the single-qubit computational basis vectors. 

Note that, unlike the unitary evolution the measurement process leads
to collapse of the state to one of the computational basis vectors
and is therefore non-unitary.

We describe below some important single qubit and two qubit quantum
gates by their actions in the computational basis.

\begin{itemize}
\bigskip{}
\item \textbf{Single qubit gates}
\end{itemize}
(A) The Quantum \textbf{NOT} (\textbf{X}) gate:

\begin{flushleft}Just like the classical NOT gate, for the quantum
NOT gate we have \[
\left|0\right\rangle \rightarrow \left|1\right\rangle \quad and\quad \left|1\right\rangle \rightarrow \left|0\right\rangle \]
\end{flushleft}

\begin{flushleft}The corresponding unitary matrix is\[
X=\left(\begin{array}{cc}
 0 & 1\\
 1 & 0\end{array}\right)\]
\end{flushleft}

\begin{flushleft}Actually there are infinitely many one qubit gates
corresponding to infinitely many 2x2 unitary matrices. We give below
two more important examples.\end{flushleft}

\bigskip{}
\begin{flushleft}(B) The \textbf{Z gate:}\end{flushleft}

\begin{flushleft}Its action in the computational basis is given by\[
\left|0\right\rangle \rightarrow \left|0\right\rangle \quad and\quad \left|1\right\rangle \rightarrow -\left|1\right\rangle \]
The corresponding unitary matrix is\[
Z=\left(\begin{array}{cc}
 1 & 0\\
 0 & -1\end{array}\right)\]
\end{flushleft}
\bigskip{}

\medskip{}
\begin{flushleft}(C) The \textbf{Hadamard} (\textbf{H}) gate:\end{flushleft}

\bigskip{}
\begin{flushleft}This is described by the \emph{Hadamard transformation}
which in the computational basis, is given by\[
\left|0\right\rangle \rightarrow \frac{1}{\sqrt{2}}(\left|0\right\rangle +\left|1\right\rangle )\quad and\quad \left|1\right\rangle \rightarrow \frac{1}{\sqrt{2}}(\left|0\right\rangle -\left|1\right\rangle )\]
The corresponding unitary matrix is\[
H=\frac{1}{\sqrt{2}}\left(\begin{array}{cc}
 1 & 1\\
 1 & -1\end{array}\right)\]
\end{flushleft}

\begin{flushleft}We generally denote the Hadamard gate by the symbol\end{flushleft}

\vspace{0.3cm}
\begin{center}\includegraphics{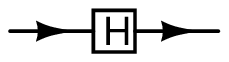}\end{center}
\vspace{0.3cm}

\bigskip{}
\begin{flushleft}Note that the Z and H gates do not have classical
analogues.\end{flushleft}

\medskip{}
\begin{flushleft}\textbf{Problem 4:} Think about a quantum $\sqrt{NOT}$
gate i.e a gate that is equivalent to a $NOT$ gate when applied twice
in succession on a qubit ( $(\sqrt{NOT})^{2}=NOT$ ). Write down it's
action on the computational basis states $\left|0\right\rangle $
and $\left|1\right\rangle $.\end{flushleft}
\medskip{}

\begin{itemize}
\newpage
\item \textbf{Two qubit gates}
\end{itemize}
(A) The \textbf{Controlled NOT (C-NOT)} gate

\begin{flushleft}This uses a control (upper) qubit and a target (lower)
qubit as inputs\end{flushleft}

\vspace{0.3cm}
\begin{center}\includegraphics{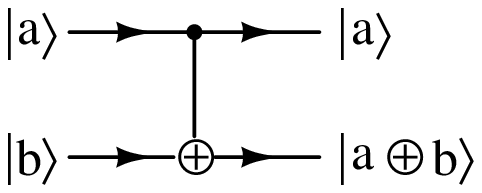}\end{center}
\vspace{0.3cm}

\begin{flushleft}The action in the computational basis is as shown
above, where $a,b\in \{0,1\}$ . The target is unchanged if the control
is off ($a=0$) and is flipped (NOTed) if the control is on ($a=1$).
The C-NOT is an important quantum gate. We illustrate below its use,
by constructing a circuit that swaps a pair of qubits in the computational
basis.\end{flushleft}

\vspace{0.3cm}
\begin{center}\includegraphics{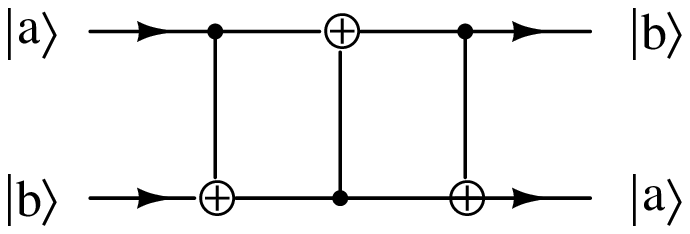}\end{center}
\vspace{0.3cm}

\begin{flushleft}The action of the circuit in the computational basis
is\[
\left|a,b\right\rangle \rightarrow \left|a,a\oplus b\right\rangle \rightarrow \left|(a\oplus b)\oplus a,a\oplus b\right\rangle =\left|b,a\oplus b\right\rangle \rightarrow \left|b,b\oplus (a\oplus b)\right\rangle =\left|b,a\right\rangle \]
\end{flushleft}

\begin{flushleft}Thus $a$ and $b$ are interchanged. Here and in
the following we are representing a two-qubit computational basis
state equivalently as\[
\left|a,b\right\rangle =\left|a\right\rangle \otimes \left|b\right\rangle =\left|a\right\rangle \left|b\right\rangle \]
\end{flushleft}
\medskip{}

If the target is initially off ($b=0$) then the C-NOT gate copies
the control qubit at the output, $\left|a,0\right\rangle \rightarrow \left|a,a\right\rangle $.

\begin{flushleft}This copying of course works for the computational
basis states. What if the control qubit is in a general state $\left|\psi \right\rangle =\alpha \left|0\right\rangle +\beta \left|1\right\rangle $
, where $\mid \alpha \mid ^{2}+\mid \beta \mid ^{2}=1$? In that case\[
\left|\psi \right\rangle \left|0\right\rangle =\alpha \left|0,0\right\rangle +\beta \left|1,0\right\rangle \rightarrow \alpha \left|0,0\right\rangle +\beta \left|1,1\right\rangle \]
\end{flushleft}

\begin{flushleft}On the other hand, if the control qubit is faithfully
copied, the final state should be\[
\left|\psi \right\rangle \left|\psi \right\rangle =\alpha ^{2}\left|0,0\right\rangle +\beta ^{2}\left|1,1\right\rangle +\alpha \beta \left|0,1\right\rangle +\beta \alpha \left|1,0\right\rangle \]
\end{flushleft}

\begin{flushleft}This is the same as above only if either $\alpha =0$
or $\beta =0$ \emph{i.e.} if the input qubit is in a computational
basis state. Would it be possible to use more complicated gates and
circuits to copy arbitrary quantum states? The answer, following from
the linearity and unitarity of quantum gates, is no . This result
goes by the name: \emph{No Cloning Theorem.}\end{flushleft}

\bigskip{}
\begin{flushleft}\textbf{Problem 5:} Design a quantum circuit that
will copy the Hadamard states $\left|\pm \right\rangle =\frac{1}{\sqrt{2}}(\left|0\right\rangle \pm \left|1\right\rangle )$
faithfully, using the quantum C-NOT gate and four Hadamard gates.
Show that this circuit is equivalent to a quantum C-NOT gate with
the control and target bits interchanged.\end{flushleft}
\bigskip{}

\bigskip{}
\begin{flushleft}\textbf{Problem 6:} Prove the \emph{No Cloning Theorem:
Any quantum copier can at best copy a set of mutually orthogonal states
and not any arbitrary unknown quantum state,} in the following way.
Consider a unitary operator $U$ that takes the product state $\left|\psi \right\rangle \left|S\right\rangle $
to $\left|\psi \right\rangle \left|\psi \right\rangle $, where $\left|\psi \right\rangle $
is the state to be copied and $\left|S\right\rangle $ is some standard
state. If this is possible for two different states $\left|\psi \right\rangle =\left|\alpha \right\rangle $
and $\left|\psi \right\rangle =\left|\beta \right\rangle $ then show
that they must be orthogonal.\end{flushleft}
\bigskip{}

\begin{flushleft}Another important two qubit gate is\end{flushleft}

\begin{flushleft}(B) The \textbf{Function} gate (\textbf{f-gate}):\end{flushleft}

\vspace{0.3cm}
\begin{center}\includegraphics{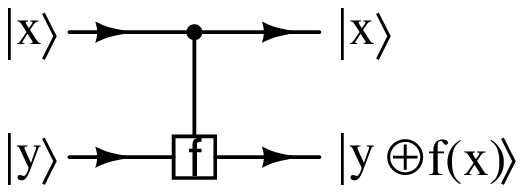}\end{center}
\vspace{0.3cm}

\begin{flushleft}The action in the computational basis is shown above.
This is a generalization of the C-NOT gate which evaluates a Boolean
function of a Boolean argument $f:\{0,1\}\rightarrow \{0,1\}$. The
C-NOT corresponds to $f(x)=x$. For $y=0$ , the action of the f-gate
is\[
\left|x,0\right\rangle \rightarrow \left|x,f(x)\right\rangle \]
\end{flushleft}

\begin{flushleft}What if the control (upper) qubit is in an arbitrary
state $\left|\psi \right\rangle =\alpha \left|0\right\rangle +\beta \left|1\right\rangle $?
In that case the result is \[
\left|\psi \right\rangle \left|0\right\rangle =\alpha \left|0,0\right\rangle +\beta \left|1,0\right\rangle \rightarrow \alpha \left|0,f(0)\right\rangle +\beta \left|1,f(1)\right\rangle \]
\end{flushleft}

\begin{flushleft}Thus the result contains the values of the function
$f$ for two possible arguments simultaneously. As we will see, this
is the key to the quantum parallelism alluded to earlier. \end{flushleft}

\bigskip{}
\begin{flushleft}\textbf{Problem 7:} Show that the quantum f-gate
is unitary, where $f:\: \{0,1\}\rightarrow \{0,1\}$ is a Boolean
function of a Boolean argument.\end{flushleft}
\bigskip{}

\bigskip{}
One may be tempted to think that the quantum gates are like probabilistic
classical gates. For example, the Hadamard gate converts the state
$\left|0\right\rangle $ to $\frac{1}{\sqrt{2}}(\left|0\right\rangle +\left|1\right\rangle ),$
which upon measurement yields the values $0$ or $1$ each with probability
$\frac{1}{2}$. This is just like a classical gate that produces the
result $0$ or $1$ each with probability $\frac{1}{2}$, depending
on say, the result of a fair coin toss. However this notion is quickly
dispelled by the observation that a second application of the Hadamard
gate will change the state back to $\left|0\right\rangle $ that yields
the value $0$ with certainty. This is of course impossible with any
classical probabilistic gate. It happens with quantum gates because
of interference between parallel channels.

More formally the Hadamard state $\frac{1}{\sqrt{2}}(\left|0\right\rangle +\left|1\right\rangle )$
is a pure state described by the density matrix $\frac{1}{2}\left(\begin{array}{cc}
 1 & 1\\
 1 & 1\end{array}\right)$ in the computational basis, whereas the output of the classical probabilistic
gate is a mixed state with the density matrix $\frac{1}{2}\left(\begin{array}{cc}
 1 & 0\\
 0 & 1\end{array}\right)$ .

\bigskip{}
\section{Bloch Sphere Representation}

Consider a general normalized one qubit state $\left|\psi \right\rangle =\alpha \left|0\right\rangle +\beta \left|1\right\rangle $.
Since \\
$\mid \alpha \mid ^{2}+\mid \beta \mid ^{2}=1$, we may parameterize\[
\alpha =\cos \frac{\theta }{2},\quad \beta =e^{i\varphi }\sin \frac{\theta }{2}\quad (0\leq \theta <\pi ,\: 0\leq \varphi <2\pi )\]

\begin{flushleft}up to an unimportant overall phase factor.\end{flushleft}

\vspace{0.3cm}
\begin{center}\includegraphics{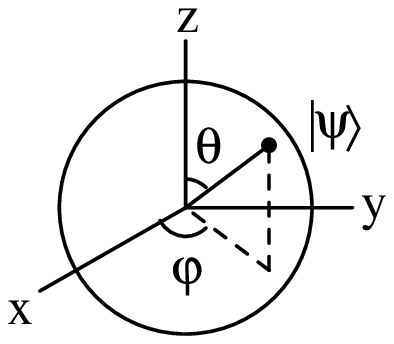}\end{center}
\vspace{0.3cm}

\begin{flushleft}The angles $\theta $ and $\varphi $ may be used
as the polar angle and azimuth of a point on the unit sphere as shown
above. This is known as the \emph{Bloch sphere representation}%
\footnote{This is the same as the \emph{Poincare sphere representation} for
the polarization states of a photon, with $\left|0\right\rangle $
representing say, the \emph{left circularly polarized} state and $\left|1\right\rangle $
representing the \emph{right circularly polarized} state. %
} of one qubit state $\left|\psi \right\rangle $ . The computational
basis states $\left|0\right\rangle $ and $\left|1\right\rangle $
are then represented by the north and south poles respectively, of
the sphere. The single qubit gates correspond to transformations on
the \emph{Bloch sphere.} For example, the X-gate (NOT) and the Z-gate
correspond to rotations through $\pi $ about the x and z axes respectively.\end{flushleft}

\bigskip{}
\begin{flushleft}\textbf{Problem 8:} \emph{Single} \textbf{}\emph{qubit
density matrix:} Show that the density matrix for a single qubit can
be expressed as\[
\rho =\frac{1}{2}(I+\overrightarrow{r}\cdot \overrightarrow{\sigma })\]
\end{flushleft}
\bigskip{}

\begin{flushleft}where $I$ is the $2\times 2$ unit matrix , $\sigma _{i}$
are the three Pauli matrices and $\: \overrightarrow{r}\: $ is an
arbitrary (radius) vector. Prove that for pure states $\: \mid \overrightarrow{r}\mid =1$
and for mixed states $\: \mid \overrightarrow{r}\mid <1$. Thus the
pure states are represented by points on the unit sphere whereas the
mixed states are represented by points inside it. This is precisely
the description we obtained above for pure states. Density matrix
formalism allows it to be extended to mixed states.\end{flushleft}

\bigskip{}
\begin{flushleft}\textbf{Problem 9:} Describe the action of the Hadamard
gate on the Bloch sphere.\end{flushleft}

\medskip{}
We conclude this section with a brief discussion about the universality
of quantum gates. It turns out that any quantum gate (and therefore
circuit) may be simulated by a combination of single qubit gates and
just the quantum C-NOT gate. There are, of course, infinitely many
single qubit gates, but fortunately they may all be obtained by combining
gates that correspond to rotations through arbitrary angles about
the y and z axes in the Bloch sphere representation. So a minimal
universal set consists of gates implementing these rotations $R_{y}(\alpha )$,
$R_{z}(\beta )$ and the quantum C-NOT gate.

\section{Classical Computation with Quantum Computers}

Any quantum computer with gates simulating the basic classical gates
can be used equally well for classical computation. We of course have
to consider the reversible classical gates, because only those have
quantum counterparts. We have seen that the Toffoli gate is a nice
classical reversible gate that can be used to simulate the basic classical
gates. So all we need is a quantum Toffoli gate. This, in terms of
its action in the computational basis, is just the classical Toffoli
gate with the input and output bits replaced by the corresponding
states.

\vspace{0.3cm}
\begin{center}\includegraphics{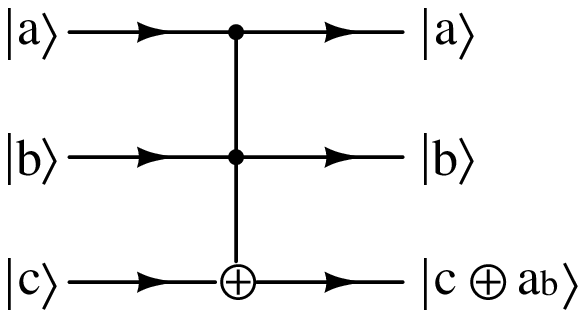}\end{center}
\vspace{0.3cm}

\begin{flushleft}Thus a quantum computer constructed in this way would
be able to do anything that a classical computer can do, equally efficiently.
If this mimicry is all that quantum computers were capable of, then
there would not be much point in discussing them. As we will see they
can do much more.\end{flushleft}

\medskip{}
\begin{flushleft}\textbf{Problem 10:} \emph{The quantum half adder:}
Using the quantum Toffoli gate and the quantum C-NOT gate construct
a quantum circuit that uses two single-qubit computational basis states
$\left|x\right\rangle $ and $\left|y\right\rangle $ as inputs and
produces the sum state $\left|x\oplus y\right\rangle $ and the carry
state $\left|xy\right\rangle $ at the output. Both input and output
may contain additional states.\end{flushleft}
\medskip{}

\section{Deutsch Problem}

Consider an arbitrary Boolean function of a Boolean argument $f:\: \{0,1\}\rightarrow \{0,1\}$.
There are, of course, four such functions corresponding to two possible
arguments and two possible values. For two of them $f(0)=f(1)$ and
these are called \emph{constant.} For the other two $f(0)\neq f(1)$
and these are called \emph{balanced.} Suppose we do not know the function,
but are given a black box or \emph{Oracle} which can evaluate it and
tell us the result. How do we decide whether the function is constant
or balanced? 

To solve this problem classically, we will have to use the oracle
twice to know its values for $0$ and $1$. David Deutsch devised
a quantum algorithm and the corresponding circuit to solve this problem
with just one call to the oracle. This in fact was the first quantum
algorithm ever written. The idea is to use a quantum oracle that in
a sense evaluates $f(0)$ and $f(1)$ simultaneously. This naturally
employs a f-gate with $f$ being our function.

\vspace{0.3cm}
\begin{center}\includegraphics{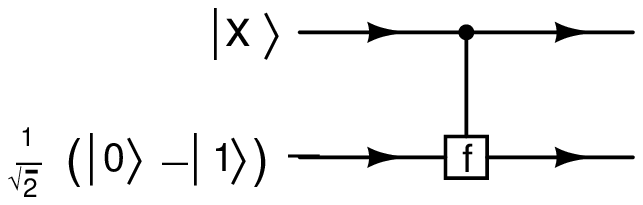}\end{center}
\vspace{0.3cm}

\begin{flushleft}In the above, the input two qubit state is\[
\left|\psi _{in}\right\rangle =\left|x\right\rangle \frac{1}{\sqrt{2}}(\left|0\right\rangle -\left|1\right\rangle )=\frac{1}{\sqrt{2}}(\left|x,0\right\rangle -\left|x,1\right\rangle )\]
\end{flushleft}

\begin{flushleft}Hence the output is $\: \left|\psi _{out}\right\rangle =\frac{1}{\sqrt{2}}(\left|x,f(x)\right\rangle -\left|x,1\oplus f(x)\right\rangle )$\end{flushleft}

\begin{flushleft}Since $f(x)=0$ or $1$, this may be written as\[
\left|\psi _{out}\right\rangle =(-1)^{f(x)}\frac{1}{\sqrt{2}}(\left|x,0\right\rangle -\left|x,1\right\rangle )=(-1)^{f(x)}\left|x\right\rangle \frac{1}{\sqrt{2}}(\left|0\right\rangle -\left|1\right\rangle )\]
\end{flushleft}

\begin{flushleft}Thus the net effect is to change the state of the
top qubit according to\[
\left|x\right\rangle \rightarrow (-1)^{f(x)}\left|x\right\rangle \]
\end{flushleft}

\begin{flushleft}\emph{i.e.} the value of the function gets kicked
back to the phase of the state $\left|x\right\rangle $.\end{flushleft}

The actual circuit used in the Deutsch's algorithm is the following.

\vspace{0.3cm}
\begin{center}\includegraphics{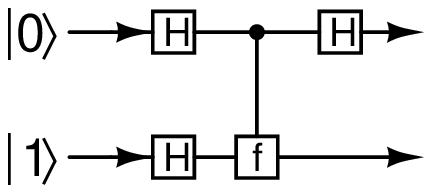}\end{center}
\vspace{0.3cm}

\begin{flushleft}The two Hadamard gates before the f-gate simply converts
the states of the input qubits according to $\left|0\right\rangle \rightarrow \frac{1}{\sqrt{2}}(\left|0\right\rangle +\left|1\right\rangle $
and $\left|1\right\rangle \rightarrow \frac{1}{\sqrt{2}}(\left|0\right\rangle -\left|1\right\rangle )$.
Now that the lower qubit is in the right state for the phase-shift
action of the f-gate, the upper qubit is transformed linearly to\[
\frac{1}{\sqrt{2}}[(-1)^{f(0)}\left|0\right\rangle +(-1)^{f(1)}\left|1\right\rangle ]\]
 \end{flushleft}

\begin{flushleft}If the function is constant \emph{i.e.} $f(0)=f(1)$
, then this is $\pm \frac{1}{\sqrt{2}}(\left|0\right\rangle +\left|1\right\rangle )$
and the final Hadamard gate the produces the state $\pm \left|0\right\rangle $.
On the other hand, if the function is balanced \emph{$\: $i.e. $f(0)\neq f(1)$
,} then the result after the f-gate is the state $\pm \frac{1}{\sqrt{2}}(\left|0\right\rangle -\left|1\right\rangle )$
for the upper qubit. Now the last Hadamard gate produces the state
$\pm \left|1\right\rangle $. Thus a single call to the quantum oracle
followed by the measurement of the upper qubit in the computational
basis, solves the problem. The speedup achieved over the classical
algorithm in this case is just a factor of $2$. However a similar
quantum algorithm for solving a generalized problem that we are going
to discuss next, would show the power of quantum computation.\end{flushleft}

\section{The Deutsch-Jozsa Algorithm}

This solves a generalization of the Deutsch problem. Let $f:\{0,1\}^{\otimes n}\rightarrow \{0,1\}$
be a Boolean function of a n-bit integer argument and assume that
we allow only those $f$ that are either \emph{constant} or yield
$0$ for exactly half of the arguments and $1$ for the rest. In the
latter case the function is called \emph{balanced.} Given an oracle
that evaluates the function for a given argument and returns the value,
the problem is to decide whether it is constant or balanced.

There are of course $2^{n}$ possible arguments corresponding to that
many different n-bit integers and to solve the problem classically,
we will have to get the function evaluated for \[
\frac{1}{2}2^{n}+1=2^{n-1}+1\]
 arguments in the worst case. This is because, with any order of evaluation,
the oracle may return $0$ (or $1$) for first half of the arguments
and we would need the value of the function for one more argument
in order to decide if it is constant or balanced. 

The computational resources required to solve the problem grows exponentially
with the (bit) size $n$ of the input \emph{i.e.} the argument, for
large $n$. In the standard terminology of computer science, such
problems are called \emph{hard.} As we will see below, the Deutsch-Jozsa
quantum algorithm is going to make this problem very \emph{easy.}

This algorithm uses a quantum f-gate that is a generalization of the
one used in the Deutsch algorithm.

\vspace{0.3cm}
\begin{center}\includegraphics{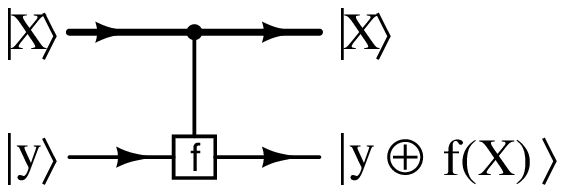}\end{center}
\vspace{0.3cm}

\begin{flushleft}The action in the computational basis is identical
with that for the ordinary f-gate, except that $\left|X\right\rangle $
here, is a computational basis state of a n-qubit register labelled
by a n-bit integer $X$. If the bottom qubit is in the Hadamard state
$\frac{1}{\sqrt{2}}(\left|0\right\rangle -\left|1\right\rangle )$
, then just as in the case of the ordinary f-gate, the state of the
upper register is transformed according to\[
\left|X\right\rangle \rightarrow (-1)^{f(X)}\left|X\right\rangle \]
\end{flushleft}

The quantum circuit used to solve the problem is the following.

\vspace{0.3cm}
\begin{center}\includegraphics{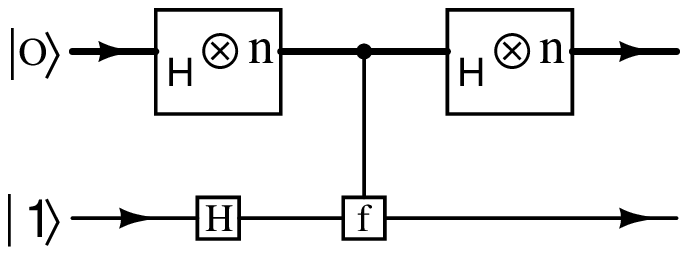}\end{center}
\vspace{0.3cm}

\begin{flushleft}Here the upper input $\left|O\right\rangle $ is
the computational basis state of the n-qubit register labelled by
the n-bit zero \emph{i.e. $\left|O\right\rangle =\left|0\right\rangle \otimes \left|0\right\rangle \otimes ........\otimes \left|0\right\rangle \quad (n\: factors)$}
and the lower input is a one-qubit computational basis state $\left|0\right\rangle $.
$\: H^{\otimes n}=H\otimes H\otimes ........\otimes H\quad (n\: factors)$
is a generalized Hadamard operator that applies the Hadamard transformation
on each of the $n$ factors of a n-qubit computational basis state
of the register.\end{flushleft}

\begin{flushleft}The effect of the first generalized Hadamard gate
on the input state of the register is\[
H^{\otimes n}\left|O\right\rangle =\frac{1}{\sqrt{2}}(\left|0\right\rangle +\left|1\right\rangle )\: \frac{1}{\sqrt{2}}(\left|0\right\rangle +\left|1\right\rangle ).......\frac{1}{\sqrt{2}}(\left|0\right\rangle +\left|1\right\rangle )=\frac{1}{\sqrt{2^{n}}}\sum _{X=0}^{2^{n}-1}\left|X\right\rangle \]
\end{flushleft}

\begin{flushleft}On the other hand the Hadamard gate acting on the
input state $\left|1\right\rangle $ of the lower qubit puts it in
the state $\frac{1}{\sqrt{2}}(\left|0\right\rangle -\left|1\right\rangle )$
which is just right for the phase-shift action of the f-gate. Thus
the f-gate changes the state of the register to\[
\frac{1}{\sqrt{2^{n}}}\sum _{X=0}^{2^{n}-1}(-1)^{f(X)}\left|X\right\rangle \]
\end{flushleft}

To see the effect of the final generalized Hadamard gate on this state,
we need to know its action on a computational basis state of the register.
This is given by\[
H^{\otimes n}\left|X\right\rangle =\frac{1}{\sqrt{2^{n}}}\sum _{Y=0}^{2^{n}-1}(-1)^{X\cdot Y}\left|Y\right\rangle \]

\begin{flushleft}where $X\cdot Y$ \emph{}is the \emph{bitwise scalar
product} defined in the following way : if $X=x_{n-1}.......x_{1}x_{0}$
and $Y=y_{n-1}.......y_{1}y_{0}$ are the binary representations of
two n-bit integers then\[
X\cdot Y=\oplus _{i=0}^{n-1}\: x_{i}y_{i}\]
\end{flushleft}

\bigskip{}
\begin{flushleft}\textbf{Problem 11:} Prove the above formula for
$H^{\otimes n}\left|X\right\rangle $.\end{flushleft}
\bigskip{}

\begin{flushleft}Thus the state of the register finally is \end{flushleft}

\[
\frac{1}{\sqrt{2^{n}}}\sum _{X=0}^{2^{n}-1}(-1)^{f(X)}H^{\otimes n}\left|X\right\rangle =\frac{1}{2^{n}}\sum _{X,Y=0}^{2^{n}-1}(-1)^{f(X)+X\cdot Y}\left|Y\right\rangle \]

\begin{flushleft}Now, the amplitude of $\left|O\right\rangle $ in
this state is $\: \frac{1}{2^{n}}\sum _{X=0}^{2^{n}-1}(-1)^{f(X)}$.\\
If $f$ is constant, then this is simply $\pm 1$. On the other hand
if $f$ is balanced, then one half of the terms in the sum precisely
cancel against the other half and the result is $0$. Hence the probability
of observing $O$ is $\, 1\, $ if $\: f\: $ is constant and is $0$
if it is balanced.\end{flushleft}

A single call to the quantum oracle followed by measurement of the
register and checking the result for $O$ , allows us to decide if
the function is constant or balanced. The quantum algorithm has achieved
an \emph{exponential speedup} over classical computation!

\medskip{}
\begin{flushleft}\textbf{Problem 12:} \emph{Bernstein-Vazirani problem}\end{flushleft}

\begin{flushleft}Given an oracle which evaluates for some $n$-bit
integer $A$, the function $f_{A}(X)=A\cdot X$ of $n$-bit integer
$X$, where $A\cdot X$ again is the bitwise scalar product, the problem
is to determine $A$. If the Deutsch-Jozsa circuit is used with the
function $f_{A}$ , then show that the final state of the $n$-qubit
register is \[
\frac{1}{2^{n}}\sum _{X,Y=0}^{2^{n}-1}(-1)^{A\cdot X+X\cdot Y}\left|Y\right\rangle =\left|A\right\rangle \]
Thus a single use of the oracle followed by measurement of the $n$-qubit
register will yield the integer $A$ with certainty.\end{flushleft}
\medskip{}

\section{Grover Search}

The problem is to search for a particular item in an \emph{unstructured}
or \emph{unsorted} database. Consider, for example, the Kolkata telephone
directory. It is, of course, arranged in the alphabetical order of
names, but not in the order of telephone numbers. Thus looking for
a particular telephone number, involves searching an unstructured
database.

It is convenient to index the database and determine the indices for
matching entries (solutions). If there are $N$ entries in the database,
then they may be indexed by integers $0,1,2.......N-1$. Assuming
that the entries occur perfectly randomly in relation to the search
field (\emph{i.e.} the telephone number in our example) , the average
number of lookups required to find a matching entry is\[
1.\frac{1}{N}+2.\frac{1}{N}+.......+N.\frac{1}{N}=\frac{1}{N}.\frac{N(N+1)}{2}=\frac{N+1}{2}\]

\begin{flushleft}If for the sake of analysis, we assume that $N=2^{n}$,
then it scales as $2^{n-1}$for large $n$. This grows exponentially
with the (bit) size $n$ of the database. Hence the problem is \emph{hard}
according to the standard definition.\end{flushleft}

Grover search is a quantum algorithm that makes the search more \emph{efficient.}
It does not make it \emph{easy} though. As we will see, the number
of lookups required to find a matching entry with high probability,
scales as $\sqrt{N}=2^{n/2}$, which is still exponential in $n$.
Note that this algorithm, unlike the Deutsch and Deutsch-Jozsa algorithms,
may not always yield the correct result. This statistical nature,
in fact, is shared by many quantum algorithms. However, given a candidate
solution, it is usually easy to check its correctness. For our problem,
one just has to look up the entry using the solution index and verify
that it contains the item being searched.

We start by defining a search function $f:\: \{0,1\}^{\otimes n}\rightarrow \{0,1\}$
such that\[
f(x)=\left\{ \begin{array}{cc}
 1 & if\: x\: is\: a\: solution\\
 0 & otherwise\end{array}\right.\]

\smallskip{}
\begin{flushleft}The search function is evaluated by a black box or
oracle which need not know the solutions beforehand. Given an index
$X$ as the argument, it just has to look up the corresponding entry
in the database and check if that contains the search item.\end{flushleft}

The oracle used in the Grover algorithm is actually a quantum oracle
that uses a generalized f-gate just as in the case of the Deutsch-Jozsa
algorithm, with the function $f$ being the search function.

\vspace{0.3cm}
\begin{center}\includegraphics{GQf.ps}\end{center}
\vspace{0.3cm}

\begin{flushleft}Its action in the computational basis is shown above.
The upper input labelled by the index $X$ , is a computational basis
state of a n-qubit index register and the lower input is a computational
basis state of an oracle qubit. The oracle qubit is flipped if and
only if, the index $X$ is a solution to the search problem.\end{flushleft}

The oracle qubit is actually initialized in the Hadamard state \\
$\frac{1}{\sqrt{2}}(\left|0\right\rangle -\left|1\right\rangle )$.
In that case the net effect is to change the state of the index register,
as in the Deutsch-Jozsa algorithm, according to\[
\left|X\right\rangle \rightarrow O\left|X\right\rangle =(-1)^{f(X)}\left|X\right\rangle \]

\begin{flushleft}where $O$ is the \emph{oracle operator.} Thus the
oracle marks the solution(s) by shifting the phase of the index state.\end{flushleft}

The complete circuit used in the Grover algorithm is shown below.

\vspace{0.3cm}
\begin{center}\includegraphics{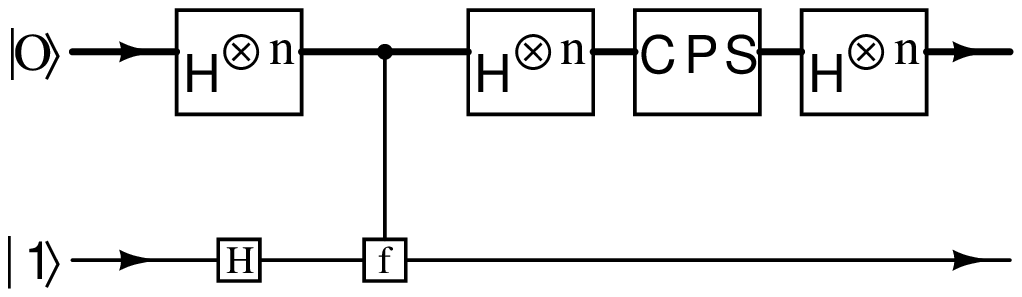}\end{center}
\vspace{0.3cm}

\begin{flushleft}The oracle qubit starts in the state $\left|1\right\rangle $,
which is then changed by the Hadamard gate to the Hadamard state $\frac{1}{\sqrt{2}}(\left|0\right\rangle -\left|1\right\rangle )$.
This is just right for the oracle operation. Similarly the index register
starts in the state $\left|O\right\rangle $ which is then transformed
by the generalized Hadamard gate $H^{\otimes n}$ to the uniform state
\[
\left|\psi \right\rangle =H^{\otimes n}\left|O\right\rangle =\frac{1}{\sqrt{N}}\sum _{X=0}^{N-1}\left|X\right\rangle \]
\end{flushleft}

which has the same amplitude for all index states $\left|X\right\rangle $.

\newpage
\begin{flushleft}Now a set of four operations is applied in the following
order.\end{flushleft}

\begin{enumerate}
\item The oracle operation $O$
\item Generalized Hadamard transformation $H^{\otimes n}$
\item Conditional phase shift (CPS): $\left|X\right\rangle \rightarrow (-1)^{1+\delta _{X,0}}\left|X\right\rangle $
. This changes the phase of all index states but $\left|O\right\rangle $
by $-1$ and is therefore, equivalent to the action of the operator
$2\left|O\right\rangle \left\langle O\right|-I$, where $I$ is the
identity operator.
\item Generalized Hadamard transformation $H^{\otimes n}$
\end{enumerate}
The product of the four is called the \emph{Grover operator} $G$.

\smallskip{}
\begin{flushleft}Note that the product of the last three is\[
H^{\otimes n}(2\left|O\right\rangle \left\langle O\right|-I)H^{\otimes n}=2\left|\psi \right\rangle \left\langle \psi \right|-I\]
\end{flushleft}

\begin{flushleft}Thus the Grover operator is $\quad G=(2\left|\psi \right\rangle \left\langle \psi \right|-I)O$\end{flushleft}

\begin{flushleft}We will see that each of the two factors in $G$
is a reflection and therefore $G$ itself is a rotation.\end{flushleft}

To show this, it is convenient to define orthonormalized states which
are uniform superpositions of the solution states (say, $M$ in number)
and the non-solution states ($N-M$ in number) separately,

\medskip{}
\begin{center}$\left|\alpha \right\rangle =\frac{1}{\sqrt{N-M}}\sum ^{''}\left|X\right\rangle $
and $\left|\beta \right\rangle =\frac{1}{\sqrt{M}}\sum ^{'}\left|X\right\rangle ,$\end{center}
\medskip{}

\begin{flushleft}where the prime and the double-prime indicate sums
over solution and non-solution states respectively. Then under the
oracle operator $O$\end{flushleft}

\smallskip{}
\begin{center}$\left|\alpha \right\rangle \rightarrow \left|\alpha \right\rangle $
but $\left|\beta \right\rangle \rightarrow -\left|\beta \right\rangle $\end{center}
\smallskip{}

\begin{flushleft}So $O$ is a reflection about $\left|\alpha \right\rangle $
in the $\left|\alpha \right\rangle ,\, \left|\beta \right\rangle $
plane. Now $\left|\psi \right\rangle $ can be expressed as \[
\left|\psi \right\rangle =\sqrt{\frac{N-M}{N}}\left|\alpha \right\rangle +\sqrt{\frac{M}{N}}\left|\beta \right\rangle =\cos \frac{\theta }{2}\left|\alpha \right\rangle +\sin \frac{\theta }{2}\left|\beta \right\rangle \]
\end{flushleft}

where $\sin \frac{\theta }{2}=\sqrt{\frac{M}{N}}$

\medskip{}
\begin{flushleft}Thus $\left|\psi \right\rangle $ is a vector in
the $\left|\alpha \right\rangle ,\, \left|\beta \right\rangle $ plane
and it is easy to see that $2\left|\psi \right\rangle \left\langle \psi \right|-I$
is a reflection about $\left|\psi \right\rangle $ in that plane.\end{flushleft}

\vspace{0.3cm}
\begin{center}\includegraphics{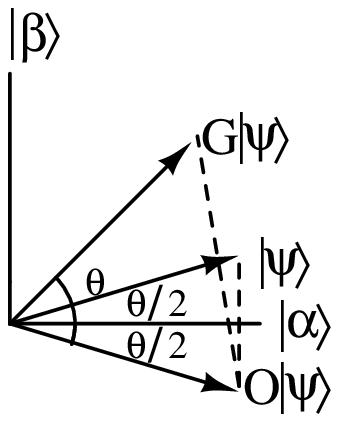}\end{center}
\vspace{0.3cm}

Simple geometry shows that the effect of these two successive reflections
on $\left|\psi \right\rangle $, is to rotate it counterclockwise
towards $\left|\beta \right\rangle $ by an angle $\theta $ in the
$\left|\alpha \right\rangle ,\, \left|\beta \right\rangle $ plane
.Thus the Grover operator changes $\left|\psi \right\rangle $ to\[
G\left|\psi \right\rangle =\cos \frac{3\theta }{2}\left|\alpha \right\rangle +\sin \frac{3\theta }{2}\left|\beta \right\rangle \]

\medskip{}
\begin{flushleft}\textbf{Problem 13:} Show algebraically that the
Grover operator $G$ rotates an arbitrary state in the $\left|\alpha \right\rangle ,\, \left|\beta \right\rangle $
plane by an angle $\theta $ in the counterclockwise direction.\end{flushleft}
\medskip{}

After $k$ iterations of the Grover sequence the state of the index
register is\[
G^{k}\left|\psi \right\rangle =\cos \frac{(2k+1)\theta }{2}\left|\alpha \right\rangle +\sin \frac{(2k+1)\theta }{2}\left|\beta \right\rangle \]

\begin{flushleft}It is clear that after sufficient number of iterations
this state would be closest to the solution space vector $\left|\beta \right\rangle $.
Then a measurement of the index register would yield a solution with
a high probability.\end{flushleft}

What is the optimum number of iterations required for this? We would
obviously need \[
\frac{(2k+1)\theta }{2}\simeq \frac{\pi }{2}\quad i.e.\quad k\simeq \frac{\pi }{2\theta }-\frac{1}{2}\]

\begin{flushleft}But for large databases \emph{i.e.} for large values
of $N$,\[
\theta =2\arcsin \sqrt{\frac{M}{N}}\simeq 2\sqrt{\frac{M}{N}}\]
\end{flushleft}

\begin{flushleft}Thus $k\simeq \frac{\pi }{4}\sqrt{\frac{N}{M}}$
. This of course depends on the number of solutions $M$. If we already
know that there is just one solution \emph{i.e.} $M=1\: $(this for
example, is true in our telephone directory example), then $k\simeq \frac{\pi }{4}\sqrt{N}$
and we need $O(\sqrt{N})$ calls to the oracle compared to $O(N)$
in the classical case. Thus the quantum algorithm achieves a \emph{quadratic
speedup} over the classical one.\end{flushleft}

\medskip{}
\begin{flushleft}\textbf{Problem 14:} Consider the Grover search for
one item ($M=1$) in a database of four entries($N=4$).What is the
optimum number of calls to the oracle required? What then is the probability
of finding the correct item?\end{flushleft}
\medskip{}

We end this section with a few comments.

\begin{itemize}
\item The Grover sequence can in principle begin with the index register
in any state in the plane spanned by $\left|\alpha \right\rangle $
and $\left|\beta \right\rangle $. We choose the uniform state $\left|\psi \right\rangle $
in this plane because, it is easily obtained from the computational
basis state $\left|O\right\rangle $ without any \emph{a priori} knowledge
about $\left|\alpha \right\rangle $ and $\left|\beta \right\rangle $.
\item We need to know the number of solutions $M$ to estimate the optimum
number of Grover iterations. This may not be known in general before
the search. Fortunately, there exist quantum algorithms for determining
the number of solutions (without actually finding them) efficiently.
\item It has been shown that the Grover search is optimal in the sense that
no other algorithm based on a quantum oracle can do better.
\end{itemize}

\section{Phase Estimation}

Suppose $U$ is a unitary operator that acts on k-qubit states and
$\left|u\right\rangle $ is an eigenstate with eigenvalue $e^{i\phi }$
:\[
U\left|u\right\rangle =e^{i\phi }\left|u\right\rangle \]

\begin{flushleft}The problem is to get the best n-bit estimate for
the phase fraction $\frac{\phi }{2\pi }$. The quantum circuit used
to solve the problem employs two new gates which we introduce in the
following.\end{flushleft}

\begin{itemize}
\newpage
\item The \textbf{Controlled-U} gate
\end{itemize}
\vspace{0.3cm}
\begin{center}\includegraphics{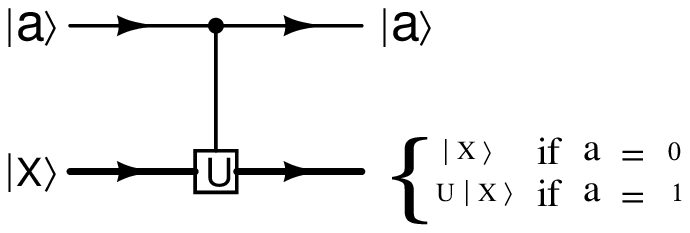}\end{center}
\vspace{0.3cm}

The action in the computational basis is shown above. The target state
$\left|X\right\rangle $ remains unmodified if the control (upper)
qubit is off ($a=0$) and a unitary operator $U$ is applied to it
if the control qubit is on ($a=1$).
\bigskip{}

\begin{itemize}
\item The \textbf{Quantum Fourier Transform (QFT)} gate
\end{itemize}

\paragraph{\textmd{Let $\alpha _{x}$ be a real number labelled by a n-bit integer
$x$ . The discrete Fourier transformation is defined by }}

\[
\alpha _{x}\: \rightarrow \: \frac{1}{\sqrt{2^{n}}}\sum _{y=0}^{2^{n}-1}(e^{2\pi i\: xy/2^{n}})\: \alpha _{y}\]

\begin{flushleft}The quantum Fourier transformation (QFT) is the quantum
analogue of the above, where the numbers $\alpha _{x}$ are replaced
by the computational basis states $\left|X\right\rangle $ of a n-qubit
register labelled by n-bit integers $X$ .\[
\left|X\right\rangle \: \rightarrow \: \frac{1}{\sqrt{2^{n}}}\sum _{Y=0}^{2^{n}-1}e^{2\pi i\: \frac{XY}{2^{n}}}\left|Y\right\rangle \]
For $n=1$ this is just our old friend the Hadamard transformation.\end{flushleft}

\smallskip{}
\begin{flushleft}The inverse transformation is \[
\left|Y\right\rangle \: \rightarrow \: \frac{1}{\sqrt{2^{n}}}\sum _{X=0}^{2^{n}-1}e^{-2\pi i\: \frac{YX}{2^{n}}}\left|X\right\rangle \]
\end{flushleft}

\bigskip{}
\begin{flushleft}\textbf{Problem 15:} Show explicitly that QFT is
a unitary transformation.\end{flushleft}
\bigskip{}

\newpage
\begin{flushleft}The gate that implements QFT is called the QFT gate
and is represented by the symbol \end{flushleft}

\vspace{0.3cm}
\begin{center}\includegraphics{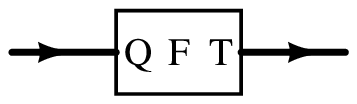} \end{center}
\vspace{0.3cm}

\begin{flushleft}The inverse transformation is implemented by running
the gate backwards.\end{flushleft}

\begin{flushleft}The actual quantum circuit used to solve the phase
estimation problem is shown below.\end{flushleft}

\vspace{0.3cm}
\begin{center}\includegraphics{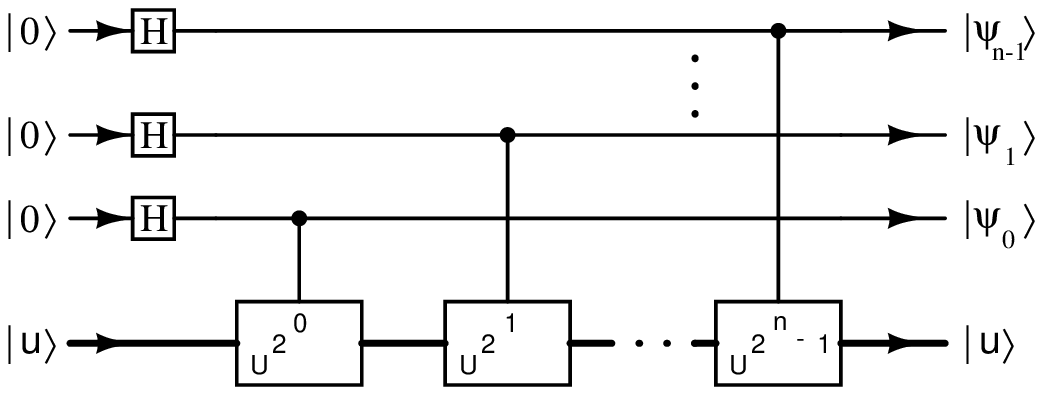}\end{center}
\vspace{0.3cm}

The target k-qubit state $\left|u\right\rangle $ is processed together
with $n$ one-qubit states (of a register) which act as controls,
through a series of controlled-$U^{2^{j}}$gates. The one-qubit states
are all initialized to $\left|0\right\rangle $ and then transformed
by Hadamard gates to the Hadamard state $\frac{1}{\sqrt{2}}(\left|0\right\rangle +\left|1\right\rangle )$
before being used as inputs in the controlled-$U^{2^{j}}$gates. Just
to see what happens at the output, let us follow through the action
of the first controlled-$U^{2^{0}}$ gate.

\[
\frac{1}{\sqrt{2}}(\left|0\right\rangle +\left|1\right\rangle )\, \left|u\right\rangle =\frac{1}{\sqrt{2}}(\left|0\right\rangle \left|u\right\rangle +\left|1\right\rangle \left|u\right\rangle )\rightarrow \frac{1}{\sqrt{2}}(\left|0\right\rangle \left|u\right\rangle +\left|1\right\rangle U^{2^{0}}\left|u\right\rangle )\]
\[
=\frac{1}{\sqrt{2}}(\left|0\right\rangle \left|u\right\rangle +e^{i2^{0}\phi }\left|1\right\rangle \left|u\right\rangle )=\frac{1}{\sqrt{2}}(\left|0\right\rangle +e^{i2^{0}\phi }\left|1\right\rangle )\, \left|u\right\rangle \]

So the net effect is to introduce a phase difference between the two
components of the control state.\[
\frac{1}{\sqrt{2}}(\left|0\right\rangle +\left|1\right\rangle )\rightarrow \frac{1}{\sqrt{2}}(\left|0\right\rangle +e^{i2^{0}\phi }\left|1\right\rangle )=\left|\psi _{0}\right\rangle \]

where $\left|\psi _{j}\right\rangle =\frac{1}{\sqrt{2}}(\left|0\right\rangle +e^{i2^{j}\phi }\left|1\right\rangle )$.

The effects of the other controlled-$U^{2^{j}}$ gates are similar.
So the final state of the control register is\[
\bigotimes _{j=n-1}^{0}\left|\psi _{j}\right\rangle =\bigotimes _{j=n-1}^{0}\frac{1}{\sqrt{2}}(\left|0\right\rangle +e^{i2^{j}\phi }\left|1\right\rangle )=\frac{1}{\sqrt{2^{n}}}\sum _{Y=0}^{2^{n}-1}e^{iY\phi }\left|Y\right\rangle \]

\medskip{}
\begin{flushleft}\textbf{Problem 16:} Prove the above relation.\end{flushleft}
\medskip{}

Suppose $\phi /2\pi =X/2^{n}$ where $X$ is some n-bit integer \emph{i.e.}
the proper fraction has exactly $n$ bits. Then the output state of
the n-qubit register is \[
\frac{1}{\sqrt{2^{n}}}\sum _{Y=0}^{2^{n}-1}e^{2\pi i\, \frac{XY}{2^{n}}}\left|Y\right\rangle \]

This is just the result of the QFT applied to the computational basis
state $\left|X\right\rangle $. Thus the inverse transformation \emph{i.e}
the QFT gate applied backwards to the output would yield $\left|X\right\rangle $.
Hence by measuring the final output in the computational basis, we
would obtain the phase fraction $\frac{\phi }{2\pi }$ exactly, with
certainty!

What if the phase fraction has more than $n$ bits? In that case we
write\[
\frac{\phi }{2\pi }=\frac{X}{2^{n}}+\epsilon \]

\begin{flushleft}where $X$, as before, is some n-bit integer and
$0<\epsilon \leq \frac{1}{2^{n+1}}$. Now the output state of the
n-qubit register in the phase estimation circuit is\[
\frac{1}{\sqrt{2^{n}}}\sum _{Y=0}^{2^{n}-1}e^{2\pi i\, \frac{XY}{2^{n}}}e^{2\pi i\, \epsilon Y}\left|Y\right\rangle \]
\end{flushleft}

\begin{flushleft}Thus the result of inverse QFT applied to this state
is\[
\frac{1}{2^{n}}\sum _{Y=0}^{2^{n}-1}e^{2\pi i\, \frac{XY}{2^{n}}}e^{2\pi i\, \epsilon Y}\sum _{Z=0}^{2^{n}-1}e^{-2\pi i\, \frac{YZ}{2^{n}}}\left|Z\right\rangle \]
\[
=\frac{1}{2^{n}}\sum _{Z=0}^{2^{n}-1}\sum _{Y=0}^{2^{n}-1}e^{2\pi i\, \frac{(X-Z)Y}{2^{n}}}e^{2\pi i\, \epsilon Y}\left|Z\right\rangle \]
\end{flushleft}

\begin{flushleft}The amplitude of $\left|X\right\rangle $ in the
above is\begin{eqnarray*}
\frac{1}{2^{n}}\sum _{Y=0}^{2^{n}-1}e^{2\pi i\, \epsilon Y} & = & \frac{1}{2^{n}}\: \frac{(e^{2\pi i\, \epsilon })^{2^{n}}-1}{e^{2\pi i\epsilon }-1}\\
 & = & \frac{1}{2^{n}}\: e^{i\pi \, \epsilon (2^{n}-1)}\: \frac{\sin \, (\pi \epsilon \, 2^{n})}{\sin \, (\pi \epsilon )}
\end{eqnarray*}
The corresponding probability is \[
P(X)=\left[\frac{1}{2^{n}}\: \frac{\sin \, (\pi \epsilon \, 2^{n})}{\sin \, (\pi \epsilon )}\right]^{2}\]
Since $0<\epsilon \, 2^{n}\leq \frac{1}{2}$, we may bound the probability
using the inequality\[
2x\leq \sin \, \pi x\leq \pi x\quad for\: \: x\, \in \, [0,\frac{1}{2}]\]
Thus $\sin \, (\pi \epsilon \, 2^{n})\geq 2\epsilon \, 2^{n}$ , $\sin \, (\pi \epsilon )\leq \pi \epsilon $
and hence\[
P(X)\geq \left(\frac{1}{2^{n}}\: \frac{2\epsilon \, 2^{n}}{\pi \epsilon }\right)^{2}=\frac{4}{\pi ^{2}}\simeq .405\]
\end{flushleft}

So a measurement of the final output state of the control register
in the computational basis, yields the first $n$ bits of the phase
fraction with a probability better than $40\%$. Actually the probability
can be made at least $1-\delta $ for any $\delta \, \varepsilon \, (0,1)$
by using $n+\log _{2}\left(2+\frac{1}{2\delta }\right)$ qubits in
the control register and rounding off the result of measurement to
$n$ bits.

\medskip{}
\begin{flushleft}\textbf{Problem 17:} Suppose we initialize the target
register in the phase estimation circuit to a linear combination $\sum c_{j}\left|u_{j}\right\rangle $
of the eigenstates $\left|u_{j}\right\rangle $ of the unitary operator
$U$, $U\left|u_{j}\right\rangle =e^{i\phi _{j}}\left|u_{j}\right\rangle $
. What do you think is obtained if the final state of the control
register is now measured in the computational basis? Assume for the
sake of simplicity that, each phase fraction $\frac{\phi _{j}}{2\pi }$
contains exactly $n$ bits.\end{flushleft}
\medskip{}

\begin{flushleft}The phase estimation algorithm has many interesting
and powerful applications. We describe next, two of its important
uses.\end{flushleft}

\section{Order and Factorization}

We begin with an elementary result in number theory. Let $N$ and
$a$ be positive integers such that $a<N$ and $a$ is \emph{coprime}
to $N$ \emph{i.e.} $gcd(a,N)=1$. Then there exists a smallest positive
integer $r<N$ such that $a^{r}=1\: mod\: N$. 

The integer $r$ is called the \emph{order} of $a\: mod\: N$. 

\medskip{}
\begin{flushleft}\textbf{Problem 18:} For a positive integer $N$
, show that the set of positive integers less than and coprime to
$N$ form a group under multiplication modulo $N$. If we denote the
order of this group by $\varphi (N)$, then $\varphi $ is called
the \emph{Euler $\varphi $-function.}\end{flushleft}
\medskip{}

\begin{flushleft}\textbf{Problem 19:} Show that the order $r$ of
$a\: mod\: N$ is a factor of $\varphi (N)$. Hence prove $a^{\varphi (N)}=1\: mod\: N$
if $a<N$ is coprime to $N$. If $p$ is a prime, then $\varphi (p)=p-1$
and we get $a^{p-1}=1\: mod\: p$ for $a<p$. Show that this is also
true when $a>p$ is not a multiple of $p$. So the general result
is that $a^{p-1}=1\: mod\: p$ if $a$ is coprime to the prime $p.$
This is known as \emph{Fermat's little theorem.} Its generalization
for arbitrary positive $N$ is due to Euler.\end{flushleft}
\medskip{}

\begin{flushleft}\textbf{Problem 20:} Find $\varphi (28)$. What is
the order of $5\: mod\: 28$?\end{flushleft}
\medskip{}

Finding the order in general, is a a hard problem in classical computation.
We describe below a special case of the phase estimation algorithm
due to Peter Shor, that makes it easy.

Suppose $N$ is a $m$-bit positive integer. For the given positive
integer $a$ less than and coprime to $N$, we define a unitary operator
$U_{a}$ such that for any $m$-qubit computational basis state $\left|X\right\rangle $\[
U_{a}\left|X\right\rangle =\left\{ \begin{array}{cc}
 \left|aX\: mod\: N\right\rangle  & if\: \: X<N\\
 \left|X\right\rangle  & otherwise\end{array}\right.\]

\medskip{}
\begin{flushleft}\textbf{Problem 21:} Show that the operator $U_{a}$
defined above is unitary\end{flushleft}
\medskip{}

It is easy to see that for each $k\, \varepsilon \, \{0,1,\: ........\: r-1\}$
the state\[
\left|u_{k}\right\rangle =\frac{1}{\sqrt{r}}\sum _{j=0}^{r-1}e^{-2\pi i\: \frac{jk}{r}}\left|a^{j}\: mod\: N\right\rangle \]
is an eigenstate of $U_{a}$ with the eigenvalue $e^{2\pi i\: \frac{k}{r}}$,
where $r<N$ is the order of $a\: mod\: N$.

\medskip{}
\begin{flushleft}\textbf{Problem 22:} Show that $U_{a}\left|u_{k}\right\rangle =e^{2\pi i\: \frac{k}{r}}\left|u_{k}\right\rangle $
for $k=0,1\: .......\: r-1$\end{flushleft}
\medskip{}

\begin{flushleft}We may in principle use the phase estimation circuit
with a $n$-qubit control register to obtain a $n$-bit estimate of
the phase fraction $\frac{k}{r}$ with a probability exceeding $40\%$
and then determine the order $r$ from that. The catch however is
that, $r$ must be known in order to prepare the target register in
the eigenstate $\left|u_{k}\right\rangle $ . The problem is obviated
by the observation that the uniform superposition of eigenstates\[
\frac{1}{\sqrt{r}}\sum _{k=0}^{r-1}\left|u_{k}\right\rangle =\left|1\right\rangle \]
and this state is therefore, easily prepared.\end{flushleft}
\medskip{}

If now the target register is initialized to the $m$-qubit computational
basis state $\left|1\right\rangle $, then a measurement of the output
state of the $n$-qubit control register would allow us to make a
$n$-bit estimate of the phase fraction $\frac{k}{r}$ for some random
$k\, \varepsilon \, \{0,1,\: ........\: r-1\}$ with a probability
of better than $40\%$.

\medskip{}
\begin{flushleft}\textbf{Problem 23:} Prove that $\frac{1}{\sqrt{r}}\sum _{k=0}^{r-1}e^{2\pi i\: \frac{jk}{r}}\left|u_{k}\right\rangle =$$\left|a^{j}\: mod\: N\right\rangle $
. Hence show that the state before the inverse QFT in the order finding
algorithm is\[
\sum _{Y=0}^{2^{n}-1}\left|Y\right\rangle U_{a}^{Y}\left|1\right\rangle =\sum _{Y=0}^{2^{n}-1}\left|Y\right\rangle \left|a^{Y}\: mod\: N\right\rangle \]
\end{flushleft}
\medskip{}

\begin{flushleft}How do we get the order $r$ from the $n$-bit estimate
of $\frac{k}{r}$ without knowing $k$? The solution hinges on two
things; the observation that $\frac{k}{r}$ is a rational number and
the following theorem from the theory of continued fractions.\end{flushleft}

\medskip{}
\begin{flushleft}\textbf{Definition:} A \emph{convergent} is a rational
fraction obtained by truncating the continued fraction expansion of
a number after a certain number of terms.\end{flushleft}

\medskip{}
\begin{flushleft}\textbf{Theorem:} If $\frac{k}{r}$ is a rational
fraction and $p$ is a positive number such that $\left|\frac{k}{r}-p\right|\leq \frac{1}{2r^{2}},$
then $\frac{k}{r}$ occurs as a convergent in the continued fraction
expansion of $p$.\end{flushleft}
\smallskip{}

\begin{flushleft}If $X$ is the result of measurement of the $n$-qubit
control register, then with better than 40\% probability, $\frac{X}{2^{n}}$
is the best $n$-bit estimate of $\frac{k}{r}$. In that case, it
differs from $\frac{k}{r}$ by at most \[
\frac{1}{2^{n+1}}+\frac{1}{2^{n+2}}+\: ........\: =\frac{1}{2^{n}}\]
Thus\[
\left|\frac{k}{r}-\frac{X}{2^{n}}\right|\leq \frac{1}{2^{n}}\]
If we choose $n>2m$ , then using $r<N$ and $N<2^{m}$ we have \[
2r^{2}<2N^{2}<2^{2m+1}\leq 2^{n}\]
Hence\[
\left|\frac{k}{r}-\frac{X}{2^{n}}\right|\leq \frac{1}{2r^{2}}\]
Thus the condition of the theorem is satisfied and $\frac{k}{r}$
can be obtained in the completely reduced form as a convergent in
the continued fraction expansion of $\frac{X}{2^{n}}$. It also turns
out that there is a unique convergent $\frac{s}{t}$ such that $t\leq r<N<2^{m}$
and $\left|\frac{s}{t}-\frac{X}{2^{n}}\right|\leq 2^{-n}$.\end{flushleft}

Once the correct convergent $\frac{s}{t}$ is determined, then assuming
that $k$ and $r$ are coprime we have, $k=s$ and $r=t$. This assumption
may be incorrect. Moreover, it is not certain that, $\frac{X}{2^{n}}$
is the best $n$-bit estimate of the phase fraction. Thus the order
$r$ obtained in this way must always be checked for $a^{r}=1\: mod\: N$
and the computation is repeated if necessary. 

How is the order related to integer factorization? Assume that we
have found the order $r$ of $a\: mod\: N$ to be even. Then \[
a^{r}=1\: mod\: N\: \Rightarrow \: (a^{\frac{r}{2}}+1)(a^{\frac{r}{2}}-1)=0\: mod\: N\]
Now, $a^{\frac{r}{2}}\neq 1\: mod\: N;$ otherwise the order of $a\: mod\: N$
would at most be $\frac{r}{2}$. Assuming further that $a^{\frac{r}{2}}\neq -1\: mod\: N$,
we conclude that one or both of $a^{\frac{r}{2}}\pm 1$ share common
factors with $N$ which may be found by evaluating $gcd(a^{\frac{r}{2}}\pm 1,N)$
by the standard Euclidean algorithm. We illustrate all these with
an example.

\medskip{}
\begin{flushleft}\textbf{Example:} \emph{Factoring $15$}\end{flushleft}

We randomly choose a number $a=7$, which is less than and coprime
to $N=15$. $15$ is a $4$ bit number \emph{i.e.} $m=4$ and $n>2m=8$.
We take $n=11$ . Suppose we find $X=1536$ to be the result of measurement
of the $11$-qubit control register in the quantum order finding circuit.
Then $\frac{X}{2^{n}}=\frac{1536}{2048}$ is (hopefully) the $11$-bit
estimate of the phase fraction $\frac{k}{r}$ for some $k\, \varepsilon \, \{0,1\: .......\: r-1\}$.
We now find the continued fraction expansion\[
\frac{1536}{2048}=\frac{1}{1+\frac{1}{3}}\]
Thus the successive convergents are $1$ and $\frac{3}{4}$, of which
$\frac{3}{4}$ is obviously the appropriate convergent. Hence assuming
$k$ and $r$ to be coprime we find $r=4$. Fortunately $r$ is even
and we check $a^{r}=7^{4}=1\: mod\: 15$. Also $a^{\frac{r}{2}}=7^{2}=4\: mod\: 15\neq -1\: mod\: 15$
. Next we find the two factors of $N=15$ by using the Euclid's algorithm
to evaluate\[
gcd(a^{\frac{r}{2}}+1,N)=gcd(50,15)=5\quad and\quad gcd(a^{\frac{r}{2}}-1,N)=gcd(48,15)=3\]
Thus $15=5\times 3$ !

\newpage
We summarize below the basic steps in the Shor factorization algorithm

\begin{enumerate}
\item Given a positive integer $N$ , we choose another positive integer
$a<N$ randomly and compute $gcd(a,N)$ using the Euclidean algorithm
on a classical computer. If $gcd(a,N)\neq 1$ , then $gcd(a,N)$ is
a factor of $N$ and we divide $N$ by it to get the other factor.
\item If $gcd(a,N)=1$ \emph{i.e. $a$} is coprime to $N$, then we find
its period $r$ modulo $N$ using the quantum order finding circuit.
\item If $r$ is even and $a^{\frac{r}{2}}\neq -1\: mod\: N$, then we compute
$gcd(a^{\frac{r}{2}}\pm 1,N)$ using the Euclidean algorithm on a
classical computer to get a pair of factors of $N.$ If not, then
we go back and repeat the steps.
\end{enumerate}
Why is quantum factorization important? It turns out that, for large
$m$ the best classical algorithm for factoring a $m$-bit integer
requires $O(e^{cm^{\frac{1}{3}}(\ln m)^{\frac{2}{3}}})$ steps in
the worst case, where $c$ is a constant. This is exponentially hard.
In fact the successes of many cryptographic algorithms, such as the
famous RSA (\emph{Rivest-Shamir-Adleman)} protocol, depend crucially
on this hardness. Shor's quantum algorithm, on the other hand achieves
the same goal in $O(m^{2+\epsilon })$ steps, where $\epsilon $ is
small, excluding the steps needed by the classical Euclidean algorithm
for gcd and the continued fraction expansion. The latter need $O(m^{3})$
steps and thus the problem becomes easy.

\section{Quantum Algorithms in General}

The phase estimation algorithm can be used to solve efficiently, many
other problems such as, period finding, discrete logarithm, Abelian
stabilizer etc efficiently using a quantum computer. Most general
among them is the

\paragraph{\textmd{\emph{Hidden subgroup problem:}} \textmd{Let $K$ be a subgroup
of a finitely generated group $G$ and $f$ be a function from $G$
to a finite set $X$, which is constant and distinct on each coset
of $K$. Given a black-box (oracle) which implements the unitary transformation
$U\left|g\right\rangle \left|h\right\rangle =\left|g\right\rangle \left|h\oplus f(g)\right\rangle $
where $\left|g\right\rangle $and $\left|h\right\rangle $ for $g\, \in \, G$,
$h\, \in \, X$, are vectors in Hilbert spaces of appropriate dimensions
and $\oplus $ is a suitable binary operation on $X$ , find a set
of generators for the hidden subgroup $K$.}}
\medskip{}

\paragraph{\textmd{All the quantum algorithms discovered so far, except the
Grover search and its generalizations, are in fact special cases of
Kitaev's algorithm for the Abelian hidden subgroup problem.}}

\section{Entanglement}

\vspace{0.3cm}
\begin{center}\includegraphics{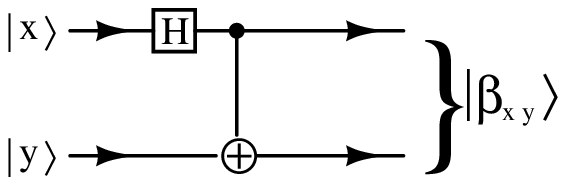}\end{center}
\vspace{0.3cm}

\begin{flushleft}In the above circuit, the output for $x=y=0$ is
the two qubit state $\left|\beta _{00}\right\rangle =\frac{1}{\sqrt{2}}(\left|00\right\rangle +\left|11\right\rangle )$.
This state can not be expressed as the product of two single-qubit
states and the two qubits are said to be \emph{entangled.} Note that,
the two qubits are correlated in this state; they are both $0$ or
$1$. This correlation survives even if the two qubits are separated
by a large distance without disturbing the state. This sort of non-local
correlation is responsible for the famous EPR (\emph{Einstein-Podolski-Rosen})
paradox. $\left|\beta _{00}\right\rangle $ and three other similarly
entangled and correlated two qubit states $\left|\beta _{01}\right\rangle =\frac{1}{\sqrt{2}}(\left|01\right\rangle +\left|10\right\rangle )$,
$\left|\beta _{10}\right\rangle =\frac{1}{\sqrt{2}}(\left|00\right\rangle -\left|11\right\rangle )$
and $\left|\beta _{11}\right\rangle =\frac{1}{\sqrt{2}}(\left|01\right\rangle -\left|10\right\rangle )$
appear in John Bell's analysis of the EPR paradox and are therefore
called Bell or EPR states.\end{flushleft}

\medskip{}
\begin{flushleft}\textbf{Problem 24:} Show that the four Bell states
form an orthonormal basis in the Hilbert space of two qubits.\end{flushleft}

\smallskip{}
\begin{flushleft}\textbf{Problem 25:} Show, for a two qubit state
which is the product of two single-qubit states, that the reduced
density matrix for each qubit in the pair corresponds to a pure state.
Calculate the reduced density matrix of the first qubit when two qubits
are in the Bell state $\left|\beta _{xy}\right\rangle $ . Does it
represent a pure state?\end{flushleft}
\medskip{}

\begin{flushleft}Far from being a nuisance, entanglement is actually
an useful resource. We describe next a couple of its applications
in quantum communication.\end{flushleft}

\section{Super-dense Coding}

Suppose Alice in Amsterdam would like to send two bits of information
to her friend Bob in Boston.%
\footnote{For alphabetical reasons the sender in this sort of scenario is always
named \textbf{A}lice, and the receiver is called \textbf{B}ob.%
} Classically of course two separate bits have to be sent. Can it be
done quantum mechanically by sending just a single qubit? The answer
is yes, provided Alice and Bob share two qubits in a Bell state, say
$\left|\beta _{00}\right\rangle $. If the two bits to be sent are
$00$ then Alice simply sends her qubit to Bob who now has the two
qubits in the state $\left|\beta _{00}\right\rangle $. If the two
bits to be sent are $01$ then Alice applies an X-gate (Quantum NOT)
to her qubit (assumed to be the first member of the pair) and sends
it to Bob who would now have the pair in the Bell state $\left|\beta _{01}\right\rangle $.
Alice similarly applies appropriate transformations to her qubit if
the bits to be sent are the other combinations $10$ or $11$ and
sends it to Bob. The general result is that the two classical bits
$xy$ are coded by a single Bell state $\left|\beta _{xy}\right\rangle $
which Bob now has. This sort of coding of a number of classical bits
by a single entangled quantum state is known as \emph{super-dense
coding.}

Since the four Bell states are orthogonal, they are certainly distinguishable
by an appropriate measurement (not necessarily in the computational
basis). Bob can therefore {}``decode'' the two qubit state which
he has and get two classical bits of information.

\medskip{}
\begin{flushleft}\textbf{Problem 26:} What transformation(s) should
Alice apply to her qubit in order to send the bit pairs $10$ and
$11$?\end{flushleft}
\medskip{}

\begin{flushleft}Super-dense coding in also useful in the detection
and correction of errors in quantum computation.\end{flushleft}

\section{Quantum Teleportation}

Alice in Amsterdam wants to send (teleport) an arbitrary one-qubit
state \\
$\left|\psi \right\rangle =\alpha \left|0\right\rangle +\beta \left|1\right\rangle $
to Bob in Boston. She can not determine the state and send the information
to Bob for its reconstruction. That would require measurements on
infinitely many copies of $\left|\psi \right\rangle $ and an arbitrary
unknown state of course can not be cloned (\emph{No Cloning Theorem}
). What can she do then? 

Assume again that Alice and Bob share an entangled EPR pair of qubits
in one of the Bell states, say $\left|\beta _{00}\right\rangle =\frac{1}{\sqrt{2}}(\left|00\right\rangle +\left|11\right\rangle )$.
The qubit to be teleported together with the EPR pair starts in the
three-qubit state\[
\left|\psi _{0}\right\rangle =\left|\psi \right\rangle \left|\beta _{00}\right\rangle =\frac{1}{\sqrt{2}}[\alpha \left|0\right\rangle (\left|00\right\rangle +\left|11\right\rangle )+\beta \left|1\right\rangle (\left|00\right\rangle +\left|11\right\rangle )]\]
 where, by convention, the first two qubits are with Alice and the
third one is with Bob.

\vspace{0.3cm}
\begin{center}\includegraphics{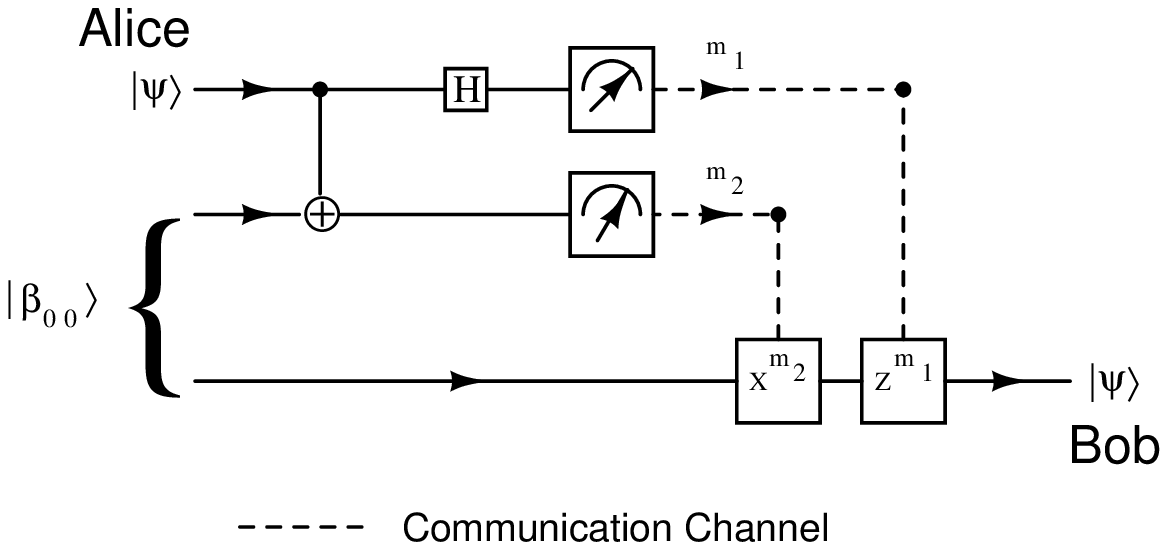}\end{center}
\vspace{0.3cm}

Now Alice puts her two qubits through a quantum C-Not gate. This entangles
the qubit to be teleported with Alice's part of the EPR pair. The
EPR pair was of course entangled to begin with. So the three qubits
end up in the entangled state\[
\left|\psi _{1}\right\rangle =\frac{1}{\sqrt{2}}[\alpha \left|0\right\rangle (\left|00\right\rangle +\left|11\right\rangle )+\beta \left|1\right\rangle (\left|10\right\rangle +\left|01\right\rangle )]\]
Next Alice sends the first qubit through a Hadamard gate and measures
her two qubits in the computational basis. The state of the three
qubits after the Hadamard gate is\begin{eqnarray*}
\left|\psi _{1}\right\rangle  & = & \frac{1}{2}[\alpha (\left|0\right\rangle +\left|1\right\rangle )(\left|00\right\rangle +\left|11\right\rangle )+\beta (\left|0\right\rangle -\left|1\right\rangle )(\left|10\right\rangle +\left|01\right\rangle )]\\
 & = & \frac{1}{2}[\left|00\right\rangle (\alpha \left|0\right\rangle +\beta \left|1\right\rangle )+\left|01\right\rangle (\alpha \left|1\right\rangle +\beta \left|0\right\rangle )\\
 &  & +\left|10\right\rangle (\alpha \left|0\right\rangle -\beta \left|1\right\rangle )+\left|11\right\rangle (\alpha \left|1\right\rangle -\beta \left|0\right\rangle )]\\
 &  & 
\end{eqnarray*}
Thus the result of Alice's measurement would be one of the pairs $00$,
$01$, $10$ or $11$ and the corresponding states in which Bob's
qubit would be left are\begin{eqnarray*}
00\rightarrow \left|\psi _{00}\right\rangle  & = & \alpha \left|0\right\rangle +\beta \left|1\right\rangle \\
01\rightarrow \left|\psi _{01}\right\rangle  & = & \alpha \left|1\right\rangle +\beta \left|0\right\rangle \\
10\rightarrow \left|\psi _{10}\right\rangle  & = & \alpha \left|0\right\rangle -\beta \left|1\right\rangle \\
11\rightarrow \left|\psi _{11}\right\rangle  & = & \alpha \left|1\right\rangle -\beta \left|0\right\rangle 
\end{eqnarray*}

If Alice now communicates her result to Bob (over a classical channel
such as telephone or e-mail or a quantum channel using super-dense
coding), he would know how to transform the state of his qubit to
$\left|\psi \right\rangle $ . If the result is $00,$ he does nothing
because his qubit is already in the state $\left|\psi \right\rangle $.
For other possible results, he has to apply an appropriate combination
of X and Z gates. If the outcome of measurement is $m_{1}m_{2},$
then the general result is\[
Z^{m_{1}}\, X^{m_{2}}\, \left|\psi _{m_{1}m_{2}}\right\rangle =\left|\psi \right\rangle \]

This seems like pure quantum magic! However it has actually been achieved
in the laboratory by teleporting a coherent photon beam, including
some deliberately introduced noise, from one room to another.%
\footnote{see \url{http://www.its.caltech.edu/~qoptics/teleport.html}%
}

We end this section with a couple of observations.

\begin{itemize}
\item Teleportation does not violate the No Cloning Theorem, as the original
state $\left|\psi \right\rangle $ to be teleported, is modified in
the process.
\item Teleportation is consistent with the principle of special relativity,
as the actual information is physically communicated at a speed necessarily
less than that of light.
\end{itemize}
\textbf{Problem 27:} Describe how a shared EPR pair in the Bell state
$\left|\beta _{11}\right\rangle $ can be used to teleport an arbitrary
single qubit state $\left|\psi \right\rangle =\alpha \left|0\right\rangle +\beta \left|1\right\rangle $.

\section{Measurement and Decoherence}

The measurement of a qubit involves its interaction with the measuring
apparatus. This could, for example, result in\begin{eqnarray*}
\left|0\right\rangle \left|m\right\rangle  & \rightarrow  & \left|0\right\rangle \left|m_{0}\right\rangle \\
\left|1\right\rangle \left|m\right\rangle  & \rightarrow  & \left|1\right\rangle \left|m_{1}\right\rangle 
\end{eqnarray*}
where $\left|m\right\rangle $ is the standard state the measuring
apparatus starts in and $\left|m_{0}\right\rangle $ and $\left|m_{1}\right\rangle $
respectively are its pointer states after the interaction, corresponding
to the qubit being in the states $\left|0\right\rangle $ and $\left|1\right\rangle $.%
\footnote{The transition may be due to the interaction Hamiltonian $H_{int}=\left|0\right\rangle \left\langle 0\right|\otimes M_{0}+\left|1\right\rangle \left\langle 1\right|\otimes M_{1}$,
where $M_{0}$ and $M_{1}$ are operators that act on the Hilbert
space of the apparatus.%
} If, however, the qubit is in an arbitrary superposition $\alpha \left|0\right\rangle +\beta \left|1\right\rangle $
, then the effect of the interaction would be\[
\left|\psi _{in}\right\rangle =(\alpha \left|0\right\rangle +\beta \left|1\right\rangle )\left|m\right\rangle \rightarrow \left|\psi _{out}\right\rangle =\alpha \left|0\right\rangle \left|m_{0}\right\rangle +\beta \left|1\right\rangle \left|m_{1}\right\rangle \]
Thus the state of the qubit gets entangled with that of the measuring
apparatus. 

\newpage
The qubit in this state is described by the reduced density matrix\begin{eqnarray*}
\rho  & = & Tr_{M}\, \left|\Psi _{out}\right\rangle \left\langle \psi _{out}\right|\\
 & = & \left|\alpha \right|^{2}\left|0\right\rangle \left\langle 0\right|+\left|\beta \right|^{2}\left|1\right\rangle \left\langle 1\right|+\alpha \beta ^{\star }\left\langle m_{1}\right.|\left.m_{0}\right\rangle \left|0\right\rangle \left\langle 1\right|+\alpha ^{\star }\beta \left\langle m_{0}\right.|\left.m_{1}\right\rangle \left|1\right\rangle \left\langle 0\right|\\
 & = & \left(\begin{array}{cc}
 \left|\alpha \right|^{2} & \alpha \beta ^{\star }\left\langle m_{1}\right.|\left.m_{0}\right\rangle \\
 \alpha ^{\star }\beta \left\langle m_{0}\right.|\left.m_{1}\right\rangle  & \left|\beta \right|^{2}\end{array}\right)
\end{eqnarray*}
where $Tr_{M}$ indicates the (partial) trace over the states of the
measuring apparatus and the states $\left|m_{0}\right\rangle $ and
$\left|m_{1}\right\rangle $ are assumed to be normalized. 

The measurement actually corresponds to the observation of the apparatus
in either of the states $\left|m_{0}\right\rangle $ or $\left|m_{1}\right\rangle $
and ideal discrimination would require $\left\langle m_{1}\right.|\left.m_{0}\right\rangle =0$
(why?). In that case the reduced density matrix is\[
\rho =\left(\begin{array}{cc}
 \left|\alpha \right|^{2} & 0\\
 0 & \left|\beta \right|^{2}\end{array}\right)\]
The vanishing of the off-diagonal elements implies complete loss of
coherence%
\footnote{Note that the loss of coherence is only apparent when the qubit is
considered in isolation. Coherence still persists in the entangled
state of the qubit and the measuring apparatus. Also, decoherence
does not explain collapse of the state vector, though the statistical
property of the reduced density matrix is the same as that of the
collapsed state.%
} between the two components of the state of the qubit. In general
$\left|\left\langle m_{1}\right.|\left.m_{0}\right\rangle \right|<1$
and the interaction with the measuring apparatus leads to reduction
in the off-diagonal elements of the reduced density matrix, implying
partial decoherence.

A qubit is really an open system interacting with its environment.
The environmental interaction decoheres the system in exactly the
same way as a measuring instrument. Decoherence during operation is
a fundamental factor limiting the reliability of quantum computation.
We illustrate its effect in the simple case of the Deutsch circuit.

\vspace{0.3cm}
\begin{center}\includegraphics{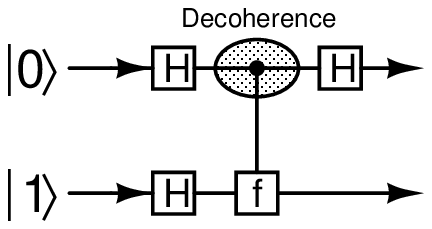}\end{center}
\vspace{0.3cm}

Suppose there is just a single decoherence interaction that acts on
the upper (control) qubit between the two Hadamard gates. In that
case the entangled state of that qubit and the environment, just before
the second Hadamard gate is\[
\frac{1}{\sqrt{2}}[(-1)^{f(0)}\, \left|0\right\rangle \left|e_{0}\right\rangle +(-1)^{f(1)}\, \left|1\right\rangle \left|e_{1}\right\rangle ]\]
where $\left|e_{0}\right\rangle $ and $\left|e_{1}\right\rangle $
are two normalized states of the environment.

The second Hadamard gate transforms this state to\begin{eqnarray*}
\left|\Psi _{out}\right\rangle  & = & \frac{1}{2}[(-1)^{f(0)}\, (\left|0\right\rangle +\left|1\right\rangle )\left|e_{0}\right\rangle +(-1)^{f(1)}(\left|0\right\rangle -\left|1\right\rangle )\left|e_{1}\right\rangle ]\\
 & = & \frac{1}{2}\left|0\right\rangle [\{(-1)^{f(0)}\left|e_{0}\right\rangle +(-1)^{f(1)}\left|e_{1}\right\rangle \}\\
 & + & \frac{1}{2}\left|1\right\rangle [\{(-1)^{f(0)}\left|e_{0}\right\rangle -(-1)^{f(1)}\left|e_{1}\right\rangle \}
\end{eqnarray*}
Assuming $\left\langle e_{0}\right.|\left.e_{1}\right\rangle $ to
be real, the reduced density matrix of the qubit in this state is%
\footnote{$Tr_{env}$ in the following, denotes the trace over the states of
the environment.%
}\begin{eqnarray*}
\rho  & = & Tr_{env}\, \left|\psi _{out}\right\rangle \left\langle \psi _{out}\right|\\
 & = & \frac{1}{2}[1+(-1)^{f(0)+f(1)}\, \left\langle e_{0}\right.|\left.e_{1}\right\rangle ]\left|0\right\rangle \left\langle 0\right|\\
 & + & \frac{1}{2}[1-(-1)^{f(0)+f(1)}\, \left\langle e_{0}\right.|\left.e_{1}\right\rangle ]\left|1\right\rangle \left\langle 1\right|
\end{eqnarray*}

\begin{flushleft}Hence the probabilities of measuring $0$ and $1$
respectively, are\begin{eqnarray*}
P_{0} & = & \frac{1}{2}[1+(-1)^{f(0)+f(1)}\, \left\langle e_{0}\right.|\left.e_{1}\right\rangle ]\\
P_{1} & = & \frac{1}{2}[1-(-1)^{f(0)+f(1)}\, \left\langle e_{0}\right.|\left.e_{1}\right\rangle ]
\end{eqnarray*}
\end{flushleft}

If the loss of coherence is complete \emph{i.e.} $\left\langle e_{0}\right.|\left.e_{1}\right\rangle =0$
, then $P_{0}=P_{1}=\frac{1}{2}$ (independent of whether $f$ is
constant or balanced) and the circuit is totally unreliable. Even
if there is only partial decoherence, the correct result is obtained
with a probability less than $1$. Hence the computation is not reliable.

When the environmental decoherence is a continuous process, the state
of the environment changes with time $t$ and the overlap is typically
$\left\langle e_{0}(t)\right.|\left.e_{1}(t)\right\rangle =e^{-\lambda t}$
. The time $\tau _{decoher}=\frac{1}{\lambda }$ is called the \emph{decoherence
time}. For $t\gg \tau _{decoher}$, decoherence is essentially complete.

\section{Devices}

Any device used for quantum computation must be able to represent
the quantum information robustly and perform a universal set of unitary
transformation corresponding to the basic quantum gates. Moreover,
we need to prepare fiducial initial states and measure the output
in an appropriate basis. The performance of a device depends primarily
on the ratio of the decoherence time $\tau _{decoher}$ to the time
scale of operation of a typical quantum gate $\tau _{op}$. We would
of course like $\tau _{decoher}$ to be as large as possible. This
is achieved by reducing the coupling with the environment. However
the manipulating devices and the measuring apparatus also couple with
the system similarly. Hence reducing the coupling too much would make
it more difficult to control the state of the device%
\footnote{This would increase $\tau _{op}$%
} and measure the result of computation. Thus a suitable compromise
has to be worked out. We describe very briefly below, some important
classes of devices that have been used for quantum computation with
some success.

\begin{itemize}
\item \textbf{Optical photon devices:} These use single photon sources and
interferometry using beam splitters, phase shifters and nonlinear
Kerr media for cross phase modulation. The qubits are represented
by the spatially different states of single photons.
\item \textbf{Cavity QED devices:} These exploit the dipole coupling of
single atoms to a few optical modes present in a high-Q cavity. The
qubits are represented by two levels of a single atom and are manipulated
using laser pulses.
\item \textbf{Ion traps:} These employ few ions cooled%
\footnote{so as to freeze their vibrational degrees of freedom%
} and trapped using electrostatic and RF electromagnetic fields. The
hyperfine levels of these ions and low lying quantized modes of vibration
of the ion chain as a whole%
\footnote{\emph{Centre of mass phonon excitations}%
} are then used to represent qubits which are manipulated by optical
laser beams.
\item \textbf{NMR:} In this case the polarized states of nuclear spins in
high magnetic fields are used to represent qubits which are manipulated
by radio frequency pulses.
\item \textbf{Quantum dots:} These are microscopic boxes created inside
metals, semiconductors and even small molecules that confine electrons
and holes by virtue of internal electrostatic fields. The quantized
energy levels of these confined charges are used to store the qubits.
The qubits are controlled by electrostatic gates (analogous to phase
shifters) and single mode wave guide couplers (analogous to beam splitters).\newpage

\end{itemize}

\section*{Acknowledgement}

First and foremost I thank Indrani Bose, but for whose enthusiasm
and friendly persistence this set of lectures would never have been
delivered and written up. I am indebted to Anjan Kundu for bringing
into focus many important and interesting issues during and outside
the lectures. I sincerely thank Ibha Chatterjee and Partha Majumdar
for their very careful reading of the draft copies of this set of
notes and for many useful comments and suggestions regarding its style,
content and presentation.


\begin{thebibliography}{1}
\bibitem{1}\textbf{Feynman Lectures on Computation:} \emph{R. P. Feynman (Edited
by Tony Hey and Robin W. Allen) , Westview Press (1999)}\\
Delightful adventure into the theory of computation and information.
Contains Feynman's vision of Quantum Computers.
\bibitem{2}\textbf{Quantum Computation and Quantum Information:} \emph{Michael
A. Nielsen and Isaac L. Chuang, Cambridge University Press (2000)
}\\
Comprehensive and readable textbook covering all aspects of Quantum
Computation and Quantum Information.
\bibitem{2}\textbf{Les Houches (1999) Lectures:} \emph{Ekert et al.}\\
Nice little introduction to Quantum Computation. Also available at
\url{http://www.qubit.org} which is the home of the Centre for Quantum
Computation at Oxford. The site contains a wealth of information and
links to active research centres all over the world.
\bibitem{3}\textbf{Lectures on Quantum Computation:} \emph{John Preskill.}\\
Available at \url{http://www.theory.caltech.edu/~preskill/ph219}
\bibitem{4}\textbf{Lectures on Quantum Computation:} \emph{N. D. Mermin.}\\
Available at \url{http://www.ccmr.cornell.edu/~mermin/qcomp/CS483.html}\end{thebibliography}
\end{document}